\documentclass[prd,showpacs,preprintnumbers,twocolumn,footinbib,amsmath,showkeys]{revtex4}
 \usepackage[dvips,final]{graphicx}
  \usepackage{amssymb,amsmath,epsfig,bm,pifont}
   \graphicspath{{./figs/}}
\usepackage{color}

\usepackage{relsize}
\newcommand{\babar}{{\mbox{\slshape B\kern-0.1em{\smaller A}\kern-0.1em
            B\kern-0.1em{\smaller A\kern-0.2em R}}}
           }
\usepackage[utf8]{inputenc}
\def\MSbar{\relax\ifmmode\overline
            {\rm MS}\else{$\overline{\rm MS}${ }}\fi}
\begin{document}
\thispagestyle{empty}
 \date{\today}
  \preprint{\hbox{RUB-TPII-01/2020}}
%\vspace*{-10mm}

\title{Pion-photon transition form factor in LCSR and tests of asymptotics
              \\}
\author{N.~G.~Stefanis}
\email{stefanis@tp2.ruhr-uni-bochum.de}
\affiliation{Ruhr-Universit\"{a}t Bochum,
             Fakult\"{a}t f\"{u}r Physik and Astronomie,
             Institut f\"{u}r Theoretische Physik II,
             D-44780 Bochum, Germany\\}
\begin{abstract}
We study the pion-photon transition form factor (TFF)
$F^{\gamma*\gamma\pi^0}(Q^2)$
using a state-of-the art implementation of light cone sum rules
(LCSRs) within fixed-order QCD perturbation theory.
The spectral density in the dispersion relation includes
all currently known radiative corrections up to the
next-to-next-to-leading-order (NNLO) and all twist contributions
up to order six.
Predictions for the TFF are obtained for various pion
distribution amplitudes (DAs) of twist two, including two-loop
evolution which accounts for heavy-quark mass thresholds.
The influence of the main theoretical uncertainties is quantified
in order to enable a more realistic comparison with the data.
The characteristics of various pion DAs are analyzed in terms of
the conformal coefficients $a_2$ and $a_4$ in comparison with the
$1\sigma$ and $2\sigma$ error regions of the data and the most
recent lattice constraints on $a_2$ with NLO and NNLO accuracy.
Our results provide more stringent bounds on the variation of the
pion DA and illuminate the corresponding asymptotic behavior of
the calculated TFF.
\end{abstract}
%PACS2010 used: 13.40.Gp Electromagnetic form factors
%               12.38.Bx Perturbative calculations
%               14.40.Be Light mesons (S=C=B=0)

\maketitle

\section{Introduction}
\label{sec:intro}
In this work we consider the pion-photon transition form factor
$F^{\gamma^*\gamma^*\pi^0}(q_1^2,q_2^2)$
for the process
$\gamma^*(q_1^2)\gamma^*(q_2^2)\rightarrow \pi^0$
with $q_1^2=-Q^2$
and $q_2^2=-q^2$
assuming $Q^2\gg q^2$
and adopting a single-tagged experimental set-up.
In that case, one measures the differential cross section
$d\sigma(Q^2,q^2=0)/dQ^2$
for the above exclusive process and selects events in which the
$\pi^0$ and one final-state electron (or positron)---the ``tag''---are
registered, while the other lepton remains undetected.

\begin{figure}[t]
\includegraphics[width=0.46\textwidth]{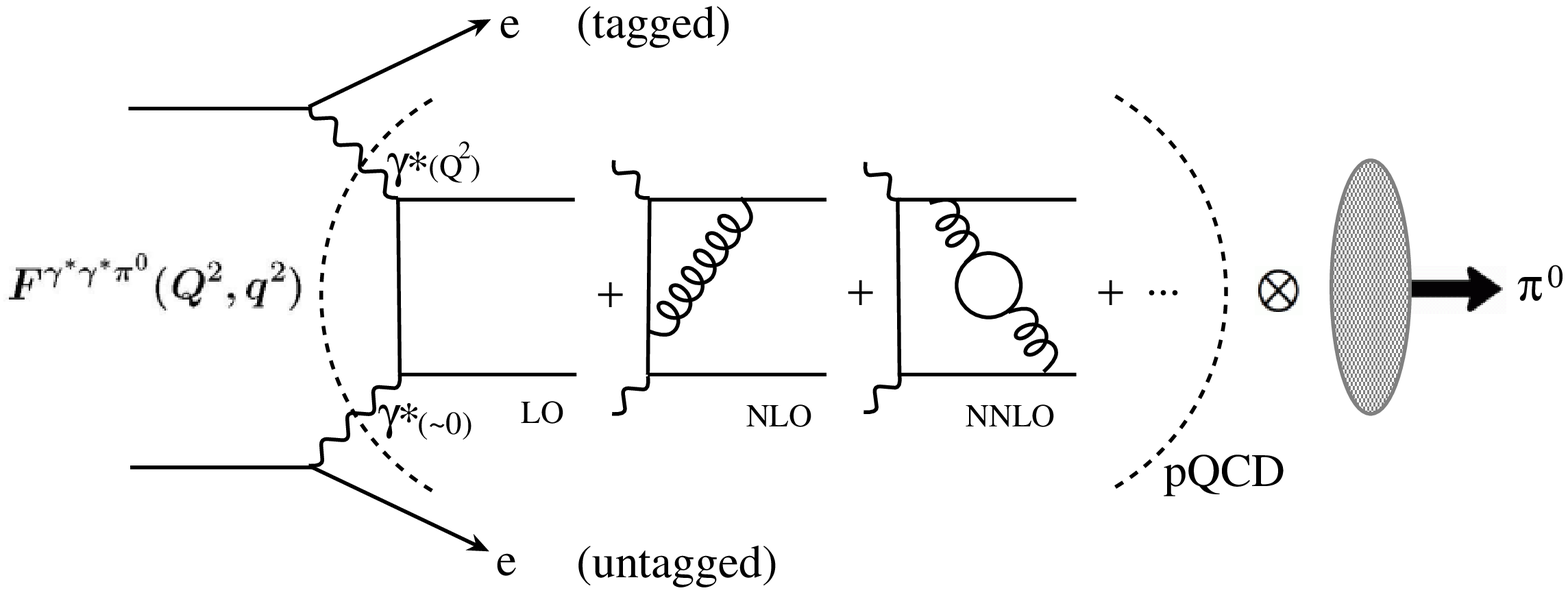}
\caption{Illustration of the single-tag $\pi^0$ production
in a two-photon process with one highly virtual photon $\gamma^*(Q^2)$
and a quasireal photon $\gamma(q^2\sim 0)$ emitted from the untagged
electron (or positron).
The TFF $\gamma^*\gamma\to q\bar{q}\to\pi^0$
is shown as the convolution of the hard quark-gluon subprocesses
within fixed-order perturbative QCD (see the text)
with the pion light-cone distribution amplitude for the pion
(shaded oval).
\label{fig:2-photon-process}
}
\end{figure}

A self-consistent calculation of the TFF within QCD encompasses various
regimes of dynamics from low $Q^2\lesssim 1$~GeV$^2$, where perturbation
theory is unreliable and nonperturbative effects are eventually more
important but poorly known, up to high $Q^2$ values where one would
expect that the perturbative contributions in terms of a power-series
expansion in the strong coupling prevail and provide an accurate
dynamical picture within perturbative QCD
(see Fig.\ \ref{fig:2-photon-process}).
There are mainly three different sources of nonperturbative
effects related to confinement that pertain to the TFF:
(i) mass generation due to Dynamical Chiral Symmetry Breaking (DCSB),
(ii) the bound-state dynamics of the pion encoded in a light-cone
parton distribution amplitude (DA),
and (iii) the hadronic content of the quasireal photon
that is emitted  from the untagged electron (or positron) at large
distances and interacts nonperturbatively with the pion.
We do not address DCSB in this work, but we refer to other
approaches which account for this and use their results in the analysis.
A reliable theoretical scheme able to include the other two
nonperturbative ingredients, together with perturbative radiative
corrections and nonperturbative higher-twist contributions, is
the method of light-cone sum rules (LCSRs)
\cite{Balitsky:1989ry,Khodjamirian:1997tk}
in combination with fixed-order perturbation theory (FOPT) within QCD.
This scheme provides computational techniques which can be used in
connection with various pion DAs and is particularly useful for the
analysis of the experimental data
\cite{Aubert:2009mc,Uehara:2012ag} that are eventually indicating
discrepant observations applying to the same phenomenon; see
\cite{Bakulev:2012nh,Stefanis:2012yw} for a detailed comparison
of various theoretical approaches and a classification scheme of the
predictions.

%%%%%%%%%%%%%%%%%%%%%%%%%%%%%%%%%%%%%%%%%%%%%%%%%%%%%%%%%%%%%%%%%%%%%%% Figure 1 - Begin
%%%%%%%%%%%%%%%%%%%%%%%%%%%%%%%%%%%%%%%%%%%%%%%%%%%%%%%%%%%%%%%%%%%%%%% Figure 1 - End
In this work we present a LCSR-based calculation of the $\pi-\gamma$
TFF which contains several new elements relative to previous
approaches: \\
(i)
The twist-two spectral density includes all presently known radiative
corrections up to the next-to-next-to-leading order (NNLO), i.e.,
up to the order of $\mathcal{O}\left(\alpha_s^2\beta_0\right)$
\cite{Mikhailov:2016klg}.
The LCSR also contains the twist-four term and the twist-six
contribution \cite{Agaev:2010aq}.
Recently, the method of LCSRs was combined with the solution of the
renormalization-group (RG) equation \cite{Ayala:2018ifo} to perform
a summation over the radiative corrections and extend its application
to momenta $Q^2<1$~GeV$^2$.
This momentum regime is outside the scope of the present
investigation. \\
(ii)
The hadronic content of the quasireal photon $\gamma(q^2)$ is included
in the LCSR by employing a physical spectral density which models the
vector-meson properties of the quasireal photon in terms of
$\rho/\omega$ resonances by means of a Breit-Wigner form. \\
(iii)
Several pion DAs are used as nonperturbative input and their
characteristics are investigated in the $(a_2,a_4)$ plane,
where $a_2$ and $a_4$ are the first nontrivial coefficients in the
conformal expansion of the twist-two pion DA.
Comparison is given with the $1\sigma$ and $2\sigma$ error regions
created within the LCSR from the combined sets of the
CELLO \cite{Behrend:1990sr},
CLEO \cite{Gronberg:1997fj},
Belle \cite{Uehara:2012ag},
and \textit{BABAR}($\leqslant 9$~GeV$^2$) \cite{Aubert:2009mc} data.
The graphical representation of the TFF predictions also includes the
recently released preliminary BESIII data
\cite{Redmer:2018uew,Ablikim:2019hff}, see also
\cite{Danilkin:2019mhd}.\footnote{The BESIII data shown in this work
were extracted from the graphics in \cite{Redmer:2018uew}
using the program Plot Digitizer.} \\
(iv)
All pion DAs used in the TFF predictions are evolved from their
normalization scale to the measured momenta using a NLO (two-loop)
evolution scheme which takes into account heavy-quark mass thresholds. \\
(v)
A crucial attenuation effect in the conformal expansion of the TFF
within the LCSR approach is worked out, which marks a crucial difference
to perturbative QCD and is of particular importance with respect to
the behavior of the TFF at large $Q^2$. \\
(vi)
To facilitate the discussion of the asymptotic characteristics of the
TFF, a new quantity is introduced which measures the scaling rate of
the scaled TFF with $Q^2$.
\\ \\
The paper is organized as follows.
In Sec.\ \ref{sec:theory} we present the theoretical formalism to
carry out the TFF calculations.
We specify the applied LCSR and discuss the important attenuation effect
related to the hadronic structure of the quasireal photon.
Our results for the pion DAs and the TFF predictions are presented in
Sec.\ \ref{sec:results}.
This section includes a dedicated discussion of the TFF asymptotics.
Our conclusions are summarized in Sec.\ \ref{sec:summary}.
The involved evolution scheme to handle the scale dependence of the
pion DA at the two-loop order by including heavy-quark thresholds is
explained in Appendix \ref{sec:global-NLO-evolution}.
Appendix \ref{sec:data-theory} completes the paper by providing a
compilation of the experimental data together with the corresponding
TFF values and uncertainties for the BMS \cite{Bakulev:2001pa} DAs.
For the first time, the analogous values for the platykurtic (pk) DA
\cite{Stefanis:2014nla} are also given.

\section{Formalism}
\label{sec:theory}

\subsection{QCD Factorization}
\label{subsec:factorization}
The amplitude $T_{\mu\nu}$ describing the process
$\gamma^{*}(q_{1})\gamma^{*}(q_{2})\rightarrow \pi^{0}(P)$
can be defined by the correlation function
\begin{eqnarray}
&& \int\! d^{4}z\,e^{-iq_{1}\cdot z}
  \langle
         \pi^0 (P)| T\{j_\mu(z) j_\nu(0)\}| 0
  \rangle
=
  i\epsilon_{\mu\nu\alpha\beta}
  q_{1}^{\alpha} q_{2}^{\beta}
\nonumber \\
&&  ~~~~~~~~~~~~~~ \times ~
  F^{\gamma^{*}\gamma^{*}\pi^0}(Q^2,q^2)\ ,
\label{eq:matrix-element}
\end{eqnarray}
%Eq (1)
where
$j_\mu=\frac{2}{3}\bar{u}\gamma_\mu u - \frac{1}{3}\bar{d}\gamma_\mu d$
is the quark electromagnetic current.
Expanding the T-product of the composite (local) current operators in
terms of $Q^2$ and $q^2$
(assuming that they are both sufficiently large),
one gets by virtue of the factorization theorem, the LO term
\cite{Lepage:1980fj,Brodsky:1981rp}
\begin{equation}
  F^{\gamma^{*}\gamma^{*}\pi}(Q^{2},q^{2})
=
  N_\text{T} %\frac{\sqrt{2}}{3}f_{\pi}
  \int_{0}^{1} dx
  \frac{1}{Q^{2}\bar{x} + q^{2}x}~\varphi_{\pi}^\text{(tw-2)}(x)
\label{eq:leading-FF-term}
\end{equation}
%Eq (2)
with $N_\text{T}=\sqrt{2}f_{\pi}/3$ and $\varphi_{\pi}^\text{(tw-2)}$ denoting
the pion DA of twist two.
For vanishing $q^2$ this expression reduces to \cite{Braun:1988qv}
\begin{eqnarray}
  \frac{3}{\sqrt{2}f_{\pi}}Q^2F_{\gamma^*\gamma\pi^0}^{(\rm LO)}(Q^2)
& = &
  \int_{0}^{1} \varphi_{\pi}^\text{(tw-2)}(x)/x
=
  \langle 1/x\rangle_{\pi}
\nonumber \\
& = &
  3(1+a_{2}+a_{4}+a_{6}+ \ldots) \, ,
\nonumber \\
\label{eq:inv-mom}
\end{eqnarray}
%Eq (3)
where we have recast the inverse moment $\langle 1/x\rangle_{\pi}$
in terms of the projection coefficients $a_n$ on the set $\{\psi_n\}$
of the eigenfunctions of the one-loop
Efremov-Radyushkin-Brodsky-Lepage (ERBL) evolution equation
\cite{Efremov:1978rn,Lepage:1980fj}:
\begin{equation}
  \varphi_{\pi}^\text{(tw-2)}(x,\mu^2)
=
  \psi_{0}(x) + \sum_{n=2,4,\ldots}^{\infty} a_{n}(\mu^2) \psi_{n}(x) \, .
\label{eq:DA-conf-exp}
\end{equation}
%Eq (4)
Here
$\psi_{0}(x)=6x(1-x)\equiv 6x\bar{x}$
is the asymptotic pion DA $\varphi_{\pi}^\text{asy}$ and the higher
eigenfunctions are given in terms of the Gegenbauer polynomials
$\psi_n(x)=6x\bar{x}C_{n}^{(3/2)}(x-\bar{x})$.

The pion DA parameterizes the matrix element
\begin{eqnarray}
  \langle 0| \bar{d}(z) \gamma_\mu\gamma_5 [z,0] u(0)
           | \pi(P)
  \rangle|_{z^{2}=0}
&& \!\!\!\!\! =
  if_\pi P_\mu \int_{0}^{1} dx e^{i x (z\cdot P)}
\nonumber \\
&& \times
  \varphi_{\pi}^\text{(tw-2)} \left(x,\mu^2\right) \, ,
\label{eq:pion-DA}
\end{eqnarray}
%Eq (5)
where the path-ordered exponential (the lightlike gauge link)
$
 [z,0]
=
 \mathcal{P}\exp \left[
                       ig \int_{0}^{z} t_{a}A_{a}^{\mu}(y)dy_{\mu}
                 \right]
$
ensures gauge invariance.
It is set equal to unity by virtue of the light-cone gauge
$z\cdot A=0$ adopted in this work.
Higher-twist DAs in the light cone operator product expansion of the
correlation function in (\ref{eq:matrix-element}) give contributions
to the TFF that are suppressed by inverse powers of $Q^2$.
Physically, $\varphi_{\pi}^\text{(tw-2)}(x,Q^2)$ describes the partition of
the pion's longitudinal momentum between its two valence partons,
i.e., the quark and the antiquark,
with longitudinal-momentum fractions
$x_q=x=(k^0+k^3)/(P^0+P^3)=k^+/P^+$
and
$x_{\bar{q}}=1-x\equiv \bar{x}$, respectively.
It is normalized to unity,
$
 \int_{0}^{1}dx \varphi_{\pi}^\text{(tw-2)}(x)=1
$, so that $a_0=1$.

The expansion coefficients $a_{n}(\mu^2)$ are hadronic parameters and
have to be determined nonperturbatively at the initial scale of
evolution $\mu^2$, but have a logarithmic $Q^2$ development via
$\alpha_s(Q^2)$ governed by the ERBL evolution equation,
see, for instance, \cite{Stefanis:1999wy} for a technical review.
The one-loop anomalous dimensions $\gamma_{n}^{(0)}$ are the
eigenvalues of $\psi_n(x)$ and are known in closed
form \cite{Lepage:1980fj}.
The ERBL evolution of the pion DA at the two-loop order is more
complicated because the matrix of the anomalous dimensions is
triangular in the $\{\psi_n(x)\}$ basis and contains off-diagonal
mixing coefficients
\cite{Dittes:1981aw,Sarmadi:1982yg,Mikhailov:1984ii,Mueller:1993hg,%%%
Mueller:1994cn,Bakulev:2002uc,Bakulev:2005vw,Agaev:2010aq}.
To obtain the TFF predictions in the present work, we employ a
two-loop evolution scheme (App.\ \ref{sec:global-NLO-evolution}),
which updates the procedure given in
Appendix D of \cite{Bakulev:2002uc}
by including the effects of crossing heavy-quark mass thresholds
in the NLO anomalous dimensions $\gamma_{n}^{(1)}$ and also
in the evolution of the strong coupling,
see, e.g., \cite{Shirkov:1994td,Bakulev:2012sm,Ayala:2014pha}.

The TFF can be expressed in more general form to read
\cite{Efremov:1978rn,Lepage:1980fj}
\begin{eqnarray}
  F_\text{QCD}^{\gamma^{*}\gamma^{*}\pi^0}\!\!\left(Q^2,q^2,\mu_{\rm F}^2\right)
= && \!\!\!\!\!
  N_\text{T} %\frac{\sqrt{2}f_{\pi}}{3}
  \int_{0}^{1} dx \,
  T\left(Q^2,q^2;\mu_{\rm F}^2;x\right)
\nonumber \\
\!\!\! && \!\!\! \times
  \varphi_{\pi}^\text{(tw-2)}\left(x,\mu_{\rm F}^2\right)
  + \mbox{h.tw.} \, ,
\label{eq:TFF-convolution}
\end{eqnarray}
%Eq (6)
where $\mu_{\rm F}$ is the factorization scale between short-distance
and large-distance dynamics and h.tw. denotes higher-twist
contributions.
The hard-scattering amplitude $T$ has a power-series expansion in terms
of the strong coupling
$a_s\equiv \alpha_{s}(\mu_{\rm R}^2)/4\pi$,
where $\mu_\text{R}$ is the renormalization scale.
In order to avoid scheme-dependent numerical coefficients, we set
$\mu_\text{F}=\mu_\text{R}\equiv \mu$
(default choice) relegating the discussion of the scheme
dependence and the factorization/renormalization scale
setting of the TFF to \cite{Stefanis:1998dg,Stefanis:2000vd}.

Then we have
\begin{equation}
  T\left(Q^2,q^2;\mu^2;x\right)
=
  T_\text{LO} + a_s~T_\text{NLO} + a_{s}^2~T_\text{NNLO} + \ldots \, ,
\label{eq:hard-scat-ampl}
\end{equation}
%Eq (7)
where the short-distance coefficients on the right-hand side can be
computed within FOPT in terms of Feynman diagrams as those depicted
in Fig.\ \ref{fig:2-photon-process}.
In our present calculation we include the following contributions,
cast in convolution form via (\ref{eq:TFF-convolution}) with
$\otimes\equiv \int_{0}^{1} dx$,
\begin{subequations}
\label{eq:T}
\begin{eqnarray}
\!  T_{\rm LO}
& \! = \! &
  T_0,
\\
\!  T_{\rm NLO}
& \! = \! & C_{\rm F}~
  T_0 \otimes \left[\mathcal{T}^{(1)}+
                      L~  V_{+}^{(0)}
              \right],
\label{eq:NLO}
\\
\!  T_{\rm NNLO}
& \! = \! &
  C_{\rm F} T_0 \otimes \!
    \left[\beta_0 T_\beta
  + T_{\Delta V}
  + T_L+  \mathcal{T}^{(2)}_c \right]\, ,
\label{eq:hard-scat-series}
\end{eqnarray}
\end{subequations}
%Eq (8a), (8b), (8c)
where the abbreviation
$
 L
\equiv
 \ln\left[\left(Q^2y+q^2\bar{y}\right)/\mu^2\right]
$ has been used \cite{Mikhailov:2009kf,Mikhailov:2016klg}.

The dominant term is \cite{Melic:2002ij,Mikhailov:2009kf}
\begin{eqnarray}
  T_\beta
\!& = &\!
  \left[\mathcal{T}_\beta^{(2)} + L\left( V_{\beta +}^{(1)} -
                                      \mathcal{T}^{(1)} \right)
                                   - \frac{L^2}{2} V^{(0)}_{+}
 \right] \, ,
\label{eq:hard-scat-nnlo-beta}
\end{eqnarray}
%Eq (9)
where
$ \beta_0
=
  \frac{11}{3}{\rm C_A} - \frac{4}3 T_{\rm R} N_f
$
is the first coefficient of the QCD $\beta$ function with
$T_{\rm R}=1/2, {\rm C_F}=4/3, {\rm C_A}=3$ for $SU(3)_c$ and
$N_f$ is the number of active flavors.

Recently, two more contributions to the NNLO radiative corrections
have been calculated in \cite{Mikhailov:2016klg} to which we refer
for their explicit expressions and further explanations.
These are
\begin{subequations}
 \label{eq:T-elements}
\begin{eqnarray}
   T_{\Delta V}
& \!\! = \!\! &
 L \Delta V^{(1)}_+ \, ,
 ~~~ \frac{V^{(1)}}{C_{\rm F}}
=
     \beta_0 V_{\beta}^{(1)} + \Delta V^{(1)}
\label{eq:hard-scat.nnlo-dv} \\
  T_L
& \!\! = \!\! &
 C_{\rm F} L \left[
           \frac{L}{2} V_{+}^{(0)}\otimes V_{+}^{(0)}
+ \mathcal{T}^{(1)}\otimes V_{+}^{(0)}
             \right]
                     \, ,
\label{eq:hard-scat-nnlo}
\end{eqnarray}
\end{subequations}
%Eq (10a), (10b)
while the term $\mathcal{T}_{c}^{(2)}$ in (\ref{eq:hard-scat-series})
has not been computed yet and is considered in this work as the main
source of theoretical uncertainties.
Finally, suffices to say that
$V_{+}^{(0)}$ and $V_{+}^{(1)}$
are the one- and two-loop ERBL evolution kernels, whereas
$V_{\beta +}^{(1)}$
is the $\beta_0$ part of the two-loop ERBL kernel,
with
$\mathcal{T}^{(1)}$
and
$\mathcal{T}_{\beta}^{(2)}$
denoting the one-loop and two-loop $\beta_0$ parts of the
hard-scattering amplitude, respectively.

\subsection{Light cone sum rules}
\label{subsec:LCSRs}
Let us now turn to the description of the TFF using a dispersion
relation within the LCSR approach.

The TFF for one highly virtual photon with the hard virtuality $Q^2$
and one photon with a small virtuality $q^2\ll Q^2$ can be expressed
in the form of a dispersion integral in the
variable $q^2 \rightarrow -s$, while $Q^2$ is kept fixed, to obtain
\begin{equation}
  F_\text{LCSR}^{\gamma^*\gamma^*\pi^0}\left(Q^2,q^2\right)
=
  N_\text{T}
  \int_{0}^{\infty}ds \frac{\rho(Q^2,s)}{q^2 + s} \, ,
\label{eq:TFF-disp-rel}
\end{equation}
%Eq (11)
where $\rho(Q^2,s)$ is the spectral density
\begin{equation}
  \rho(Q^2,s)
=
    \rho^\text{h}(Q^2,s)\theta(s_0-s)
  + \rho^\text{pert}(Q^2,s)\theta(s-s_0) \, .
\label{eq:rho-phen}
\end{equation}
%Eq (12)
The first term $\rho^\text{h}(Q^2,s)$ models the hadronic (h) content
of the spectral density,
\begin{equation}
  \rho^\text{h}(Q^2,s)
=
  \sqrt{2}f_\rho F^{\gamma^*\rho\pi}(Q^2)\delta(s-m_{\rho}^2) \, ,
\label{eq:hadr-spec-dens}
\end{equation}
%Eq (13)
while $\rho^\text{pert}(Q^2,s)$ denotes the
QCD part in terms of quarks and gluons,
calculable within perturbative QCD,
\begin{eqnarray}
  \rho^{\text{pert}}(Q^{2},s)
& = &
  \frac{1}{\pi} \text{Im} F_\text{QCD}^{\gamma^*\gamma^*\pi^0}(Q^2, -s, -i\epsilon )
\nonumber \\
& = &
   \rho_{\text{tw-2}}
  +\rho_{\text{tw-4}}
  +\rho_{\text{tw-6}}
  +\ldots\, .
\label{eq:rho-twists}
\end{eqnarray}
%Eq (14)
Each of these terms can be computed from the convolution of the
associated hard part with the corresponding DA of the same twist
\cite{Khodjamirian:1997tk}.
Below some effective hadronic threshold in the vector-meson channel,
the photon emitted at large distances is replaced in
$F^{\gamma^{*}V\pi^0}$ by a vector meson $V=\rho$, $\omega$, etc.,
using for the corresponding spectral density a phenomenological
ansatz, for instance, a $\delta$-function model.

Thus, after performing the Borel transformation
$1/(s+q^2)\rightarrow \exp\left(-s/M^2\right)$,
with $M^2$ being the Borel parameter,
one obtains the following LCSR
(see \cite{Agaev:2010aq,Mikhailov:2009kf,Mikhailov:2016klg} for
more detailed expositions)
\begin{widetext}
\begin{eqnarray}
&&  Q^2 F_\text{LCSR}^{\gamma*\gamma*\pi^0}\left(Q^2,q^2\right)
=
  N_\text{T}f_\pi
  \left[
        \frac{Q^2}{m_{\rho}^2+q^2}
        \int_{x_{0}}^{1}
        \exp\left(
                  \frac{m_{\rho}^2-Q^2\bar{x}/x}{M^2}
            \right)
  \bar{\rho}(Q^2,x)
  \frac{dx}{x}
  + \! \int_{0}^{x_0} \bar{\rho}(Q^2,x)
        \frac{Q^2dx}{\bar{x}Q^2+xq^2}
  \right] \, ,
\label{eq:LCSR-FQq}
\end{eqnarray}
%Eq (15)
\end{widetext}
where the spectral density is given by
\begin{equation}
  \bar{\rho}(Q^2,s)
=
  (Q^2+s) \rho^{\text{pert}}(Q^2,s) \, .
\label{eq:rho-bar}
\end{equation}
%Eq (16)
For simplicity, we have shown above the LCSR expression for the simple
$\delta$-function model to include the $\rho$-meson resonance into
the spectral density.
However, the actual calculation of the TFF predictions to be
presented below, employs a more realistic Breit-Wigner form, as
suggested in \cite{Khodjamirian:1997tk} and used in
\cite{Mikhailov:2009kf}.
This reads
\begin{equation}
  \delta(s-m_\text{V}^2)
\longrightarrow
  \Delta_\text{V}(s)
\equiv
  \frac{1}{\pi}
  \frac{m_\text{V} \Gamma_\text{V}}{(m_\text{V}^2 - s)^2
  + m_\text{V}^2 \Gamma_\text{V}^2} \, ,
\label{eq:Breit-Wigner}
\end{equation}
%Eq (17)
where the masses and widths of the $\rho$ and $\omega$ vector mesons
are given by
$m_\rho=0.770$~GeV,
$m_\omega=0.7826$~GeV,
$\Gamma_\rho=0.1502$~GeV, and
$\Gamma_\omega=0.00844$~GeV, respectively.
The other parameters entering (\ref{eq:LCSR-FQq}) are
$s =\bar{x}Q^2/x$ with $\bar{x}\equiv 1-x$,
$x_0 = Q^2/\left(Q^2+s_0\right)$, and the effective threshold in the
vector channel is $s_0\simeq 1.5$~GeV$^2$.
The stability of the LCSR is ensured for values of the Borel parameter
$M^2$ varying in the interval $M^2\in [0.7-1.0]$~GeV$^2$
\cite{Bakulev:2011rp,Bakulev:2012nh,Stefanis:2012yw,Mikhailov:2016klg}.
By allowing a stronger variation towards larger values
$M^2\in[0.7 - 1.5]$~GeV$^2$
\cite{Agaev:2010aq,Agaev:2012tm}, the TFF prediction receives an
uncertainty of the order $[-1.6 - 7.2]\%$ \cite{Mikhailov:2016klg}
that becomes negligible at large $Q^2$.

Note at this point that the LCSR in (\ref{eq:LCSR-FQq}) includes in an
effective way the nonperturbative long-distance properties of the
real photon in terms of the duality interval $s_0$ and the
masses of the vector mesons that are absent in the pQCD formulation
of the TFF, but play an important role in the kinematic region
$Q^2\lesssim s_0$ and $x_0\lesssim 0.5$ (cf.\ the first term in Eq.\
(\ref{eq:LCSR-FQq})).
The real-photon limit $q^2 \to 0$ can be taken
in (\ref{eq:LCSR-FQq}) by simple substitution because there are no
massless resonances in the vector-meson channel.
Thus, this equation correctly reproduces the behavior of the TFF for a
highly virtual and a quasireal photon from the asymptotic limit
$Q^2\rightarrow \infty$ down to the hadronic normalization scale of
$Q^2\sim 1$~GeV$^2$, as measured in single-tag experiments.
For still lower momenta, outside the validity range of the standard
LCSR scheme, other approaches may be more preferable
\cite{Ayala:2018ifo,Hoferichter:2018dmo,Hoferichter:2018kwz,%
Eichmann:2019tjk,Roig:2014uja,Guevara:2018rhj,Masjuan:2012wy}.
The pertinent role of subleading power corrections to the TFF has
been investigated in \cite{Wang:2017ijn,Shen:2019vdc}.

Using the conformal expansion for $\rho_\text{tw-2}$, the spectral
density can be expressed in the form
\begin{eqnarray}
  \bar{\rho}\left(Q^2,x\right)
= && \!\!\!
  \sum_{n=0,2,4,\ldots}a_{n}\left(Q^2\right)
    \bar{\rho}_{n}\left(Q^2,x\right)
  + \bar{\rho}_\text{tw-4}\left(Q^2,x\right)
\nonumber \\
&& + \bar{\rho}_\text{tw-6}\left(Q^2,x\right)
+ \ldots \, ,
\label{eq:rho-bar-15}
\end{eqnarray}
%Eq (18)
where
\begin{eqnarray}
  \bar{\rho}_{n}\left(Q^2,x\right)
\!\!\!&\!\!=\!\! &\!\!\!
    \bar{\rho}_{n}^{(0)}(x)
\!\! +\!a_{s}\bar{\rho}_{n}^{(1)}(Q^2,x)
\!\!+\!a_{s}^{2}\bar{\rho}_{n}^{(2)}(Q^2,x)
    + \ldots, \nonumber \\
\bar{\rho}_{n}^{(0)}(x)&=& \psi_n(x);~\, a_s= a_{s}(Q^2) \, ,
\label{eq:rho-n}
\end{eqnarray}
%Eq (19)
with the elements $\bar{\rho}_{n}^{(i)}$ being given in Appendix B
of Ref.\ \cite{Mikhailov:2016klg}.

The dispersive analysis here includes the twist-four and
twist-six spectral densities in explicit form.
The $\bar{\rho}_\text{tw-4}$ spectral density is given by
\begin{equation}
  \bar{\rho}_{\text{tw-4}}(Q^2,x)
=
  \frac{\delta^2_\text{tw-4}(Q^2)}{Q^2}
  x\frac{d}{dx}\varphi^\text{(tw-4)}(x)\Bigg|_{x=Q^2/(Q^2+s)}\, ,
\label{eq:rho-tw-4}
\end{equation}
%Eq (20)
where the twist-four coupling parameter takes values in the range
$
 \delta^2_\text{tw-4}(\mu^{2}=1~\rm{GeV}^2)
\approx
 \lambda_{q}^{2}/2
=
 0.19\pm 0.04$~GeV$^2
$
and is closely related to the average virtuality $\lambda_{q}^{2}$
of vacuum quarks
\cite{Mikhailov:1986be,Mikhailov:1988nz,Mikhailiov:1989mk,%%%
Bakulev:1991ps,Mikhailov:1991pt},
defined by
$
 \lambda_{q}^{2}
\equiv
 \langle
        \bar{q}(ig\sigma_{\mu\nu}G^{\mu\nu})q
 \rangle/(2\langle \bar{q}q\rangle)
=
 0.4 \pm 0.05~\text{GeV}^2
$.
Details on its estimation and evolution can be found in
\cite{Bakulev:2002uc},
whereas the sensitivity of the TFF to its variation was examined
in \cite{Bakulev:2003cs}.
In the present analysis the evolution of $\delta^2_\text{tw-4}$
is also included.
Expression (\ref{eq:rho-tw-4}) is evaluated with the asymptotic form of
the twist-four pion DA \cite{Khodjamirian:1997tk}
\begin{equation}
  \varphi_{\pi}^\text{(tw-4)}(x,\mu^2)
= \frac{80}{3} \delta^2_\text{tw-4}(\mu^2) x^2(1-x)^2 \, ,
\label{eq:tw-4-DA}
\end{equation}
%Eq (21)
while more complicated renormalon-inspired forms were considered
in \cite{Agaev:2005rc,Bakulev:2005cp} confirming that at the $1\sigma$
error level the data processing is virtually unchanged so that
Eq.\ (\ref{eq:tw-4-DA}) is sufficient.
The twist-six part of the spectral density, i.e.,
$\bar{\rho}_{\text{tw-6}}(Q^{2},x)
=
 (Q^2+s)\rho_{\text{tw-6}}(Q^2,s)
$,
was first derived in \cite{Agaev:2010aq}.
An independent term-by-term calculation in \cite{Mikhailov:2016klg}
confirmed this result.
We quote it here in the form
\begin{widetext}
\begin{eqnarray}
    \bar{\rho}_\text{tw-6}(Q^{2}\!,x)
=
    8\pi \frac{C_\text{F}}{N_c}
    \frac{ \alpha_s\langle\bar{q} q\rangle^2}{f_\pi^2}\frac{x}{Q^4}
    \left[
        \!-\!
        \left[\frac{1}{1-x}\right]_+
        \!+\!\left(2\delta(\bar{x})-4 x\right)\!+\!
        \left(
         3x+2x\log{x}
        \!+\!
        2x\log{}\bar{x}
        \right)
    \right] \, ,
\label{eq:tw-6}
\end{eqnarray}
\end{widetext}
%Eq (22)
where the plus prescription
$
 [f(x,y)]_{+}
 =
 f(x,y) - \delta(x-y)\int_{0}^{1}f(z,x) dz
$
is involved,
while
$\alpha_s=0.5$
and
$
  \langle \bar{q} q\rangle^2
=
 \left(0.242 \pm 0.01 \right)^6$ GeV$^6$
at the scale $\mu^2=1$~GeV$^2$ \cite{Gelhausen:2013wia}.

To obtain detailed numerical results for the TFF
$\mathcal{F}(Q^2)$
using (\ref{eq:LCSR-FQq}),
we employ several DAs from different approaches with
various shapes encoded in their conformal coefficients $a_n$.
The latter are determined at their native normalization scale
(as quoted in the referenced approaches)
by means of the moments of the pion DA
\begin{equation}
  \langle \xi^{N} \rangle_{\pi}
\equiv
  \int_{0}^{1} \varphi_{\pi}^\text{(tw-2)}(x,\mu^2) (x-\bar{x})^{N}dx \, ,
\label{eq:moments}
\end{equation}
%Eq (23)
where $\xi = x - \bar{x}$ and $N=2,4, \ldots$.
The expansion coefficients $a_n$ can be expressed in terms of the
moments
$\langle\xi^N\rangle_\pi$
as follows
\begin{widetext}
\begin{eqnarray}
  a_{2n}
=
  \frac{2}{3}
  \frac{4n+3}{(2n+1)(2n+2)2^{2n}}
  \sum_{m=0}^{n} (-1)^{(n-m)}
  \frac{\Gamma(2n+2m+2)}{\Gamma(n+m+1)\Gamma(n-m+1)\Gamma(2m+1)}
  \langle \xi^{2m} \rangle_{\pi} ~~~~~ (n=0,1,2,3 \ldots) \, .
\label{eq:a-n-vs-xi-n}
\end{eqnarray}
\end{widetext}
%Eq (24)

\subsection{Attenuation effect}
\label{subsec:attenuation}
We now turn our attention to an effect that marks a crucial difference
between the perturbative approach and the use of a dispersion relation
and has important consequences for the scaling behavior of the TFF.

As shown in \cite{Mikhailov:2016klg} (see Fig.\ 1 and Eq.\ (20) there),
and discussed here further, the leading-twist expression for a given
harmonic $n$ of the TFF within the LCSR scheme is not the same as in
pQCD.
The TFF considered in this work is
\begin{eqnarray}
  Q^2F^{\gamma^*\gamma\pi^0}(Q^2)
&\!\! = \!\!&
  \left[
        F_0(Q^2) + \sum_n a_n(Q^2) F_n(Q^2)
  \right] \nonumber \\
&&  + F^\text{tw-4}(Q^2) + F^\text{tw-6}(Q^2) \, ,
\label{eq:TFF-all}
\end{eqnarray}
%Eq (25)
where the terms in the square brackets represent the twist-two
contribution.
While the result based on factorization is given by the inverse moment
of $\bar{x}$ with respect to $\psi_n(x)$
\begin{equation}
  Q^2 F_{n}^\text{pQCD}(Q^2)
\! = \!
  \int_{0}^{1} \psi_n(x) \frac{dx}{\bar{x}}
= 3 \, ,
\label{eq:F-n-pQCD}
\end{equation}
%Eq (26)
the analogous LCSR expression deviates from that because of the
hadronic structure of the quasireal photon taken into account via the
vector meson dominance.
At the leading twist-two level, Eq.\ (\ref{eq:rho-bar-15}) reduces to the
Born approximation
\begin{eqnarray}
  Q^2 F_{n}^\text{LCSR}(Q^2)
&\!  = \! &
  \frac{Q^2}{m_\rho^{2}}
  {\rm e}^{\frac{m_{\rho}^2}{M^2}}
  \int_{x_{0}}^1 {\rm e}^{\frac{-Q^2\bar{x}}{M^2x}} \psi_n(x)
  \frac{dx}{x} \nonumber \\
&&  + \int_{0}^{x_0} \psi_n(x) \frac{dx}{\bar{x}} \, ,
\label{eq:F-n-LCSR}
\end{eqnarray}
%Eq (27)
where the involved parameters are defined below
Eq.\ (\ref{eq:Breit-Wigner}) and the spectral density is given by
Eq.\ (\ref{eq:rho-bar-15}).

\begin{figure*}[t]
\centering
\includegraphics[width=0.45\textwidth]{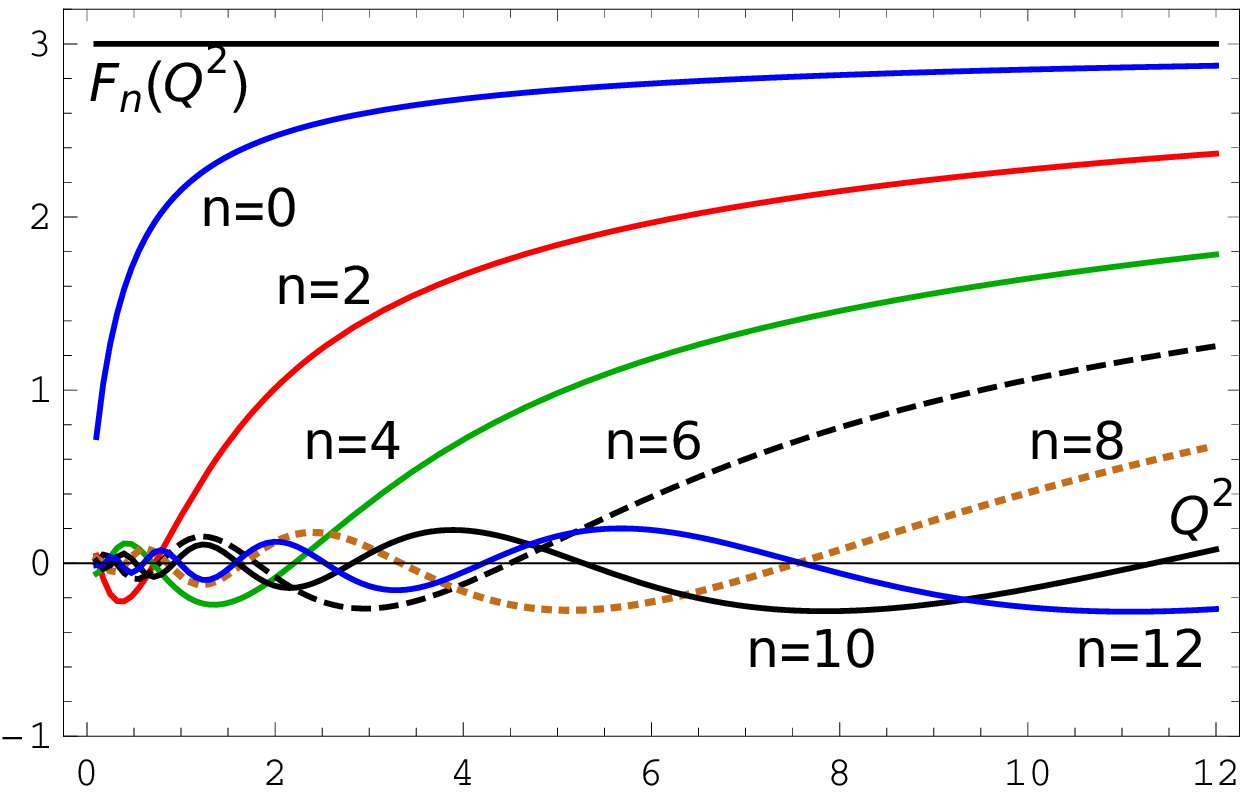}
\hfill
\includegraphics[width=0.436\textwidth]{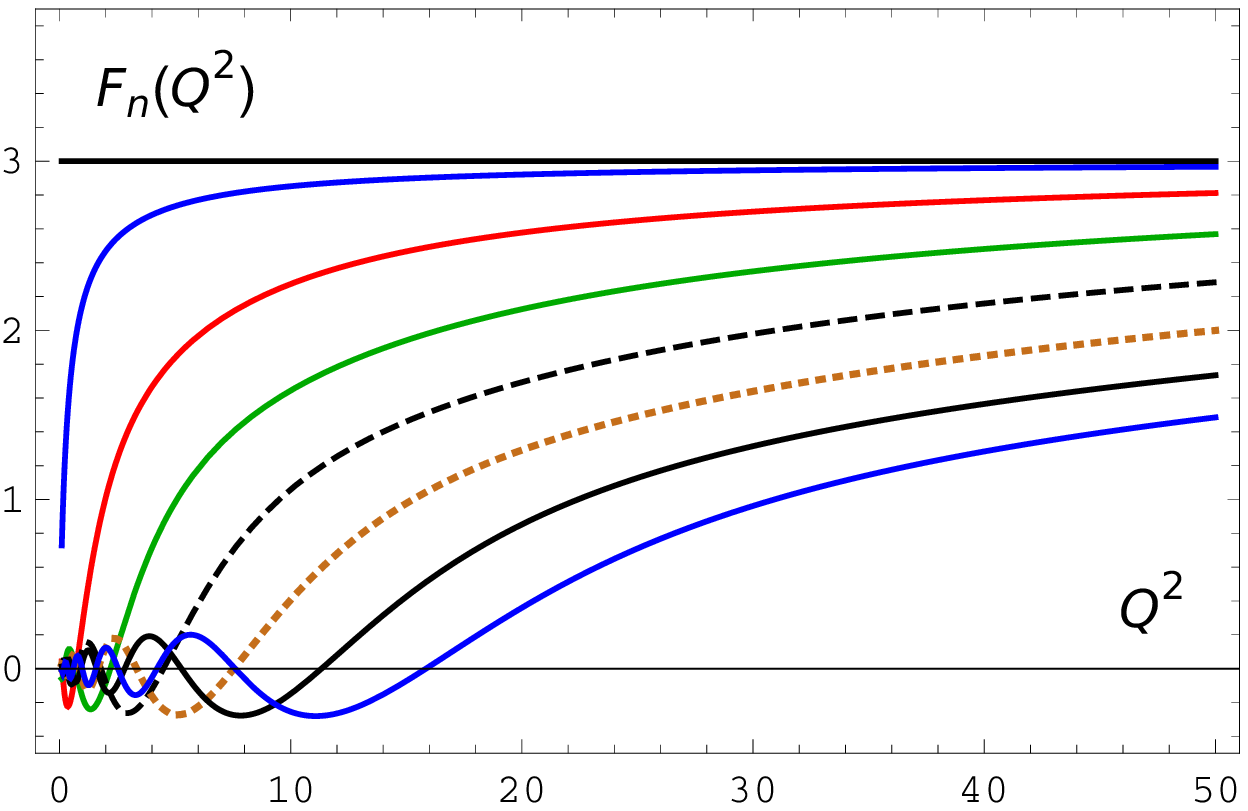}
\caption{Attenuation effect of the TFF due to the variation of the
partial terms of the scaled transition form
factor $Q^2F_{n}^\text{LCSR}(Q^2)$ with $Q^2$,
cf.\ Eq.\ (\ref{eq:F-n-LCSR}).
The left panel shows the domain below 12~GeV$^2$, while the right
panel covers the large $Q^2$ regime up to 50~GeV$^2$, where higher
harmonics shown up to $n=12$ give sizeable contributions because they
do not oscillate.
\label{fig:harmonics}}
\end{figure*}

As long as $Q^2\gg s_0$ and
$x_0=\left(1+s_0/Q^2\right)^{-1}\rightarrow 1$,
the hadronic part of the quasireal photon in the spectral density
$\bar{\rho}(Q^2,x)$ is suppressed at large $Q^2$.
As a result, all harmonics contribute at once like in perturbative QCD
and $Q^2F_{n}^\text{LCSR}(Q^2) \rightarrow 3$
on account of $6\int_0^1dx xC_n^{(3/2)}(x-\bar{x})=3$.
However, for $Q^2 \sim \mathcal{O}(s_0)$, both terms in
Eq.\ (\ref{eq:F-n-LCSR}) contribute with comparable magnitudes.
Thus, nonperturbative higher-twist contributions controlled by $s_0$
in Eq.\ (\ref{eq:F-n-LCSR}) are no more suppressed.
At the same time, the vector-meson generated factor entails an
attenuation effect of the conformal expansion
so that the $\psi_n$ harmonics contribute
successively in pace with $Q^2$, see Fig.\ \ref{fig:harmonics}.
The upshot of this nonperturbative attenuation effect is that
the LCSR-based TFF predictions can deviate significantly from
those obtained in perturbative QCD.

%%%%%%%%%%%%%%%%%%%%%%%%%%%%%%%%%%%%%%%%%%%%%%%%%%%%%%%%%%%%%%%%%%%%%%% Figure 2 - Begin
%%%%%%%%%%%%%%%%%%%%%%%%%%%%%%%%%%%%%%%%%%%%%%%%%%%%%%%%%%%%%%%%%%%%%%% Figure 2 - End

The core observations from the graphics in Fig.\ \ref{fig:harmonics}
are the following:
(i) The strongest contribution to the form factor stems from the
zeroth order term $F_0(Q^2)$ which grows uniformly.
(ii) The higher partial terms oscillate with zero crossings clustering
at $Q^2\lesssim 1$~GeV$^2$.
These oscillations cause an attenuation effect in the sense that
harmonics of higher order start to contribute at larger and larger
$Q^2$ values one following the other as $n$ grows.
For instance, $\psi_{12}$ starts to grow uniformly only beyond
the zero crossing around 15~GeV$^2$ (right panel).
(iii) Therefore, the more harmonics with positive coefficients are
included in the conformal expansion of the pion DA, the stronger the
attenuated enhancement of the form factor becomes as $Q^2$ grows.
(iv) The decrease of the conformal coefficients due to ERBL evolution
is only logarithmic and is thus insufficient to compensate for this
enhancement, though at asymptotically large $Q^2$ values it
finally prevails.
(v) These considerations apply not only to DAs with a large number
of expansion coefficients, they are also valid to a less degree for
DAs with a few number of coefficients but having an inverse hierarchy
\cite{Agaev:2012tm}.
(vi) On the other hand, at $Q^2<1$~GeV$^2$ only the term $F_0(Q^2)$
contributes so that, irrespective of how many conformal coefficients
are included in the pion DA representation, the total form factor will
be dominated by the $\psi_0$ harmonic.

\section{Analysis of the results}
\label{sec:results}

In this section we present our results for the pion DA and the TFF
predictions calculated within the LCSR approach described in the
previous section.

\subsection{Pion DAs}
\label{subsec:pi-DAs}
To obtain predictions for the TFF, we evaluate (\ref{eq:rho-bar-15})
in the LCSR (\ref{eq:LCSR-FQq}) using for the physical
spectral density in (\ref{eq:hadr-spec-dens}) the Breit-Wigner form
(\ref{eq:Breit-Wigner}) and employing various conformal coefficients
$a_n$ at their normalization scale.
We consider two such scales $\mu_1=1$~GeV and $\mu_2=2$~GeV, depending
on the particular pion DA.
If $\mu_2$ is not the native normalization scale, the ERBL evolution
scheme discussed in App.\ \ref{sec:global-NLO-evolution} is applied
and numerical results for the TFF predictions with the BMS and the pk
DAs are given.
This scheme, presented here for the first time, works for any
polynomial order of the conformal expansion and accounts for the
crossing of heavy-quark flavors at the two-loop level.
Though numerically the impact on the TFF predictions is relatively small,
the increased accuracy suffices to improve the scaling behavior of the TFF
at large $Q^2$, as one can see by comparing the BMS results with those
in \cite{Bakulev:2011rp}.

The coefficients $a_2, a_4, a_6$ of various models for the pion DA
are given in Table \ref{tab:pion-DAs-a6} at both scales $\mu_1$ and
$\mu_2$.
Using these values, one can readily compute the corresponding
moments (\ref{eq:moments}) using Eq.\ (\ref{eq:a-n-vs-xi-n}).
This table also includes the values of the inverse moment at the scale
$\mu_2$.
Because broad, concave distributions cannot be adequately represented
in terms of only the lowest three coefficients, the corresponding
inverse moments of the DSE-DB, DSE-RL, and the holographic AdS/QCD DAs
are calculated within the $\alpha_{-}$ representation
given by Eq.\ (\ref{eq:DSE-DA}) \cite{Chang:2013pq,Raya:2015gva}.
Here the abbreviation DSE means Dyson-Schwinger equations with the
label DB referring to the use of the most advanced Bethe-Salpeter
kernel while RL denotes the rainbow ladder approximation.

%%%%%%%%%%%%%%%%%%%%%%%%%%%%%%%%%%%%%%%%%%%%%%%%%%%%%%%%%%%%%%%%%%%%%%% Table I - Begin
\begin{center}
\begin{table*}[th]
\caption{Conformal coefficients $a_2$, $a_4$, $a_6$ for various pion
DAs discussed in the text at two typical normalization momentum scales
$\mu_1 = 1$~GeV and
$\mu_2=2$~GeV.
If $\mu_2$ is not the initial scale, NLO ERBL evolution in the global
scheme is employed, see App.\ \ref{sec:global-NLO-evolution}.
The range of the BMS and platykurtic DAs is related to the determination
of $a_2$ and $a_4$ from QCD sum rules with nonlocal condensates using
$\lambda_{q}^2=0.40$~GeV$^2$ and $\lambda_{q}^2=0.45$~GeV$^2$,
respectively.
They cause the variation of the TFF predictions shown in the form of
a green shaded band in
Fig.\ \ref{fig:linear-scaled-TFF}.
The coefficient $a_2$ of the CZ DA was originally given at
the scale $\mu=0.5$~GeV: $a_{2}^\text{CZ}=2/3$ \cite{Chernyak:1983ej}.
For the extrapolation to higher scales see \cite{Bakulev:2002uc}.
Higher conformal coefficients up to and including $a_{12}$ for the
DSE-DB and DSE-RL DAs at the scale $\mu_2$ can be found in
\cite{Raya:2015gva}.
The coefficients up to and including $a_{20}$ at the
scale $\mu_1$ of the holographic AdS/QCD DA
$\varphi_{\pi}^\text{hol}(x)=(8/\pi)\sqrt{x\bar{x}}$
are tabulated in \cite{Brodsky:2011yv}.
They were calculated here by means of the expression
$
 \left\langle\xi^{2n}\right\rangle_{\pi}^\text{AdS/QCD}
=
 \frac{1}{4}\frac{B\left(3/2,(2n+1)/2\right)}{B(3/2,3/2)}
$
[$B(x,y)$ being the Euler Beta function]
in combination with Eq.\ (\ref{eq:a-n-vs-xi-n}).
The lattice results of \cite{Bali:2019dqc} with NNLO (two loops) and
NLO (one loop) matching to the \MSbar scheme are quoted separately,
where the subscript $r$ denotes the systematic uncertainty due to the
nonperturbative renormalization.
They were obtained from a combined extrapolation to the chiral and
continuum limit with associated uncertainties labeled by the subscripts
$m$ and $a$, respectively.
The statistical errors of the data after extrapolation are given in
sub- and super-scrip form.
The question mark (?) in the lattice result of \cite{Braun:2015axa}
indicates that it was not extrapolated to the continuum limit.
}
\begin{ruledtabular}
\begin{tabular}{lccccccc}
Pion DA                                      & $a_2(\mu_1)$          & $a_4(\mu_1)$         & $a_6(\mu_1)$    & $a_2(\mu_2)$    & $a_4(\mu_2)$   & $a_6(\mu_2)$      & $\langle1/x\rangle_\pi(\mu_2)$
\\\hline \hline
BMS \cite{Bakulev:2001pa, Mikhailov:2016klg} \ding{54} & $0.203_{-0.057}^{+0.069}$  & $-0.143_{-0.087}^{+0.094}$  & 0 & $0.149_{-0.043}^{+0.052}$ & $-0.096_{-0.058}^{+0.063}$ & 0 & $3.16^{+0.09}_{-0.09}$
\\
BMS range                                     & $[0.146, 0.272]$      & $[-0.23, -0.049]$    & 0         & $[0.11, 0.20]$            & $[-0.15, -0.03]$               & 0 & --
\\
platykurtic \cite{Stefanis:2014nla} \ding{60} & $0.0812_{-0.025}^{+0.0345}$  & $-0.0191_{-0.0287}^{+0.0337}$ & 0 & $0.057^{+0.024}_{-0.019}$ & $-0.013^{+0.022}_{-0.019}$                    & 0                            & $3.13^{+0.14}_{-0.10}$
\\
platykurtic range                             & $[0.0562, 0.1156]$           & $[-0.0478, 0.0147]$          & 0 & $[0.04, 0.08]$            & $[-0.03, 0.01]$                & 0 & --
\\
DSE-DB \cite{Chang:2013pq,Raya:2015gva} \ding{115} & --                      & --                           & --                             & 0.149     & 0.076           & 0.031        & 4.6
\\
DSE-RL \cite{Chang:2013pq,Raya:2015gva} $\bigtriangledown$  & --           & --        & --        & 0.233  & 0.112        & 0.066        & 5.5
\\
AdS/QCD \cite{Brodsky:2011yv} $\bigtriangleup$     & $7/48$ & $11/192$     & $5^3/2^{12}$     & 0.107       & 0.038        & 0.0183       & 4.0
\\
Light-Front QM \cite{Choi:2014ifm} $\bigcirc$      & 0.0514                & -0.0340              & -0.0261          & 0.035              & $-0.0227$       &-0.0153       & 2.99
\\
NL$\chi$ QM \cite{Nam:2006sx} $\square$            & 0.0534                & -0.0609              & -0.0260          & 0.037              & $-0.041$        &-0.015        & 3.18
\\
CZ (this work) \ding{110}                    & 0.56                  & 0                    & 0                & 0.412              & 0               & 0            & 4.24
\\
Lattice \cite{Braun:2015axa}                 & --                    & --                   & --               & 0.1364(154)(145)(?)& --              & --           & --
\\
Lattice (NNLO) \cite{Bali:2019dqc}           & --                    & --                   & --               &$0.101^{+17}_{-17}(12)_r(10)_a(5)_m$  & --           & -- & --
\\
Lattice (NLO) \cite{Bali:2019dqc}            & --                    & --                   & --               &$0.078^{+18}_{-19}(16)_r(13)_a(5)_m$  & --           & -- & --
\\
\end{tabular}
\end{ruledtabular}
\label{tab:pion-DAs-a6}
\end{table*}
\end{center}
%%%%%%%%%%%%%%%%%%%%%%%%%%%%%%%%%%%%%%%%%%%%%%%%%%%%%%%%%%%%%%%%%%%%%%% Table I - End

\begin{figure}[t]
\includegraphics[width=0.45\textwidth]{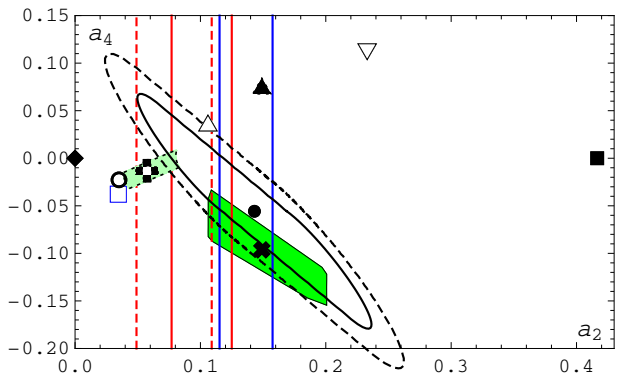}
\caption{Various pion DAs in terms of their conformal coefficients
$(a_2,a_4)$ at the scale $\mu_2=2$~GeV (Table \ref{tab:pion-DAs-a6})
shown in comparison with the
$1\sigma$ (solid line) and $2\sigma$ (dashed line) error regions
created from the combined analysis of the
CELLO \cite{Behrend:1990sr},
CLEO \cite{Gronberg:1997fj},
Belle \cite{Uehara:2012ag},
and \textit{BABAR}($\leqslant 9$~GeV$^2$) \cite{Aubert:2009mc} data
within LCSRs with the central point of the nonlinear fit marked by
the symbol $\bullet$.
The larger rectangle shows the range of values for BMS-like DAs from
\cite{Bakulev:2001pa} and the smaller one the analogous region for
the platykurtic DAs derived in
\cite{Stefanis:2014nla,Stefanis:2015qha}.
The vertical lines mark the constraints on $a_2$ from two lattice
determinations: \cite{Bali:2019dqc} NLO (dashed red lines) and
NNLO (solid red lines);
\cite{Braun:2015axa} (solid blue lines).
The asymptotic DA is denoted by \ding{117}.
The other designations are given in Table \ref{tab:pion-DAs-a6}
with further explanations in the text.
\label{fig:a2-a4-pion-DAs}
}
\end{figure}

The graphical representation of the pion DAs is displayed in Fig.\
\ref{fig:a2-a4-pion-DAs} at the scale $\mu_2=2$~GeV in terms of
the DA projections on the plane $(a_2,a_4)$
using the symbols given in Table \ref{tab:pion-DAs-a6}.
The experimental constraints are expressed in the form of $1\sigma$
(solid line) and $2\sigma$ (dashed line) error regions generated
from the combined analysis of the
CELLO \cite{Behrend:1990sr},
CLEO \cite{Gronberg:1997fj},
Belle \cite{Uehara:2012ag},
and \textit{BABAR}($\leqslant 9$~GeV$^2$) \cite{Aubert:2009mc} data
within LCSRs, see \cite{Bakulev:2002uc,Mikhailov:2016klg,Schmedding:1999ap} for
further explanations.
The two slanted rectangles represent the constraints imposed by the QCD
sum rules with nonlocal condensates used in \cite{Bakulev:2001pa}
in connection with the determination of the
Bakulev-Mikhailov-Stefanis (BMS) DAs.
The larger one corresponds to the average vacuum quark virtuality
$\lambda_{q}^{2}(\mu^2\approx 1~\mbox{GeV}^2)=0.4$~GeV$^2$,
whereas the smaller rectangle was determined in
\cite{Stefanis:2014nla,Stefanis:2015qha}
using the slightly larger but still admissible value
$\lambda_{q}^{2}(\mu^2\approx 1~\mbox{GeV}^2)=0.45$~GeV$^2$
(see \cite{Bakulev:2002hk} and references cited therein).
It contains pion DAs with a characteristic platykurtic profile
\cite{Stefanis:2014nla} (see Sec.\ \ref{subsec:pk-DA}).

This figure also contains the lattice constraints on $a_2$ at the scale
$\mu_2$ from \cite{Bali:2019dqc} (red vertical lines further to
the left) as well as those from \cite{Braun:2015axa} (blue vertical
lines).
The presented intervals in both cases are calculated by combining
errors in quadrature.
The results at the scale $\mu_2$ are
$a_2=0.101_{-0.024}^{+0.024}$ (NNLO) and
$a_2=0.078_{-0.029}^{+0.031}$ (NLO) \cite{Bali:2019dqc},
whereas $a_2=0.136\pm 0.021$ \cite{Braun:2015axa}.
A linear combination of errors would slightly overestimate the
combined uncertainties yielding somewhat larger intervals
of $a_2$ values.
The results from \cite{Bali:2019dqc}, quoted in
Table \ref{tab:pion-DAs-a6}, were obtained from a combined chiral
and continuum limit extrapolation at the
NNLO and NLO level.
This treatment differs from that applied in \cite{Braun:2015axa},
where no extrapolation to the continuum limit was carried out.
This is indicated in Table \ref{tab:pion-DAs-a6} by the question
mark (?).
The chiral extrapolation was included in the first parenthesis together
with the statistical error, while the renormalization error is given in
the second parenthesis.
The general tendency of the new lattice estimates seems to favor DAs
with a smaller value of $a_2$.
Recall that the second moment $\langle \xi^2\rangle_\pi$
(or equivalently $a_2$) gives information only on the variance
statistic
$\sigma^2[\varphi]=\frac{1}{4}\langle \xi^2\rangle_\pi$
of the pion DA and contains no information about its shape
in the central region.
To this end, one needs the kurtosis statistic
$
 \beta_2[\varphi]
=
 \frac{\langle \xi^4\rangle_\pi}{\left(\langle \xi^2\rangle_\pi\right)^2}
$
which measures the peakedness or flatness of a distribution
in terms of the fourth moment, see \cite{Stefanis:2015qha} for a
quantitative discussion.

%%%%%%%%%%%%%%%%%%%%%%%%%%%%%%%%%%%%%%%%%%%%%%%%%%%%%%%%%%%%%%%%%%%%%%% Figure 3 - Begin
%%%%%%%%%%%%%%%%%%%%%%%%%%%%%%%%%%%%%%%%%%%%%%%%%%%%%%%%%%%%%%%%%%%%%%% Figure 3 - End

\subsection{Platykurtic pion DA}
\label{subsec:pk-DA}
The particularity of the platykurtic pion DA \cite{Stefanis:2014nla}
derives from the fact that it contextually and mathematically
encapsulates in its profile the two consequences of confinement:
(i) the appearance of nonlocal vacuum expectation values whose
expansion in terms of local operators involves the virtuality
$\lambda_{q}^2$ of the vacuum quarks \cite{Mikhailov:1986be} and
(ii) dynamical chiral symmetry breaking (DCHB) and mass generation
\cite{roberts2020reflections}.
The first feature entails quark correlations at a \emph{finite}
distance $1/\lambda_q \sim 0.3$ fermi, while the second one entails
the mass dressing of the confined quark propagator, see, for instance,
\cite{Roberts:2016vyn}.
As argued in \cite{Stefanis:2014nla,Stefanis:2015qha}, these effects
induce distinctive geometrical characteristics of the pion DA.
While the first one leads to the suppression of the endpoint regions
$x=0,1$, the second one enhances the central region around $x=1/2$.

The net result of this competition is a unimodal distribution with a
unique short-tailed platykurtic profile.
Take away the quark correlations and the endpoints of the DA get too
strong resulting into a broad concave distribution in the whole $x$
range similar to that obtained with DSE \cite{Raya:2015gva}.
Leave aside the mass dressing and the central region is more or less
depleted giving rise to a bimodal distribution whose bimodality
strength is controlled by the nonlocality parameter $\lambda_{q}^{2}$.
When $\lambda_{q}^{2}=0$, one gets an infinite correlation length
corresponding to the use of local condensates in the QCD sum rules.
Such a situation gives rise to the Chernyak-Zhitnitsky (CZ)
pion DA \cite{Chernyak:1983ej}.
In contrast to bimodal DAs, like CZ and BMS, the platykurtic DA
yields in the middle point
$\varphi_{\pi/\text{pk}}^\text{(tw-2)}(x=1/2,\mu_1)=1.264$
in agreement with the LCSR calculation \cite{Braun:1988qv}
$\varphi_{\pi}^\text{(tw-2)}(x=1/2,\mu_1)=1.2\pm 0.3$.
For the derivation of the platykurtic DA and its range
(Table \ref{tab:pion-DAs-a6} and small (green) strip in
Fig.\ \ref{fig:a2-a4-pion-DAs}), we refer to
\cite{Stefanis:2014nla,Stefanis:2015qha}.

From this figure we make the following striking observations:
(i) The platykurtic strip shows a positive correlation between the
coefficients $a_2$ and $a_4$, while the BMS-type DAs (larger green
rectangle) have coefficients with a negative correlation between
them.
(ii) Also the arrangement of the $1(2)\sigma$ error regions exhibits
an anticorrelation pattern between $a_2$ and $a_4$.
(iii) Nevertheless, the platykurtic strip overlaps with the data regions
at its upper right corner where the coefficients are given by
\begin{equation}
  a_2(\mu_2) \approx 0.08, ~~~ a_4(\mu_2) \approx -0.009
\label{eq:best-pk}
\end{equation}
%Eq (28)
corresponding to the moments
\begin{equation}
  \langle \xi^2 \rangle_\pi
  \approx 0.229, ~~~
  \langle \xi^4 \rangle_\pi
  \approx 0.106 \, .
\label{eq:mom2-4-best-pk}
\end{equation}
%Eq (29)
(iv) Remarkably, just there it also enters the range of the NNLO lattice
constraints on $a_2$ from \cite{Bali:2019dqc}, while it mostly overlaps
with the analogous NLO region.
The observed agreement extends to the values of the second moment
$\langle \xi^2 \rangle_\pi$.
One has from \cite{Bali:2019dqc}
\begin{subequations}
\begin{eqnarray}
  \langle \xi^2 \rangle_\pi^\text{NNLO}(\mu_2)
& = &
  0.234^{+6}_{-6}(4)_r(4)_a(2)_m, \\
  \langle \xi^2 \rangle_\pi^\text{NLO}(\mu_2)
& = &
  0.106^{+6}_{-6}(5)_r(5)_a(2)_m
\label{eq:mom2-lattice}
\end{eqnarray}
\end{subequations}
%Eq (30a) (30b)
which gives after adding the errors in quadrature
the values
$0.234\pm 0.0085$ and $0.227\pm 0.0095$, respectively.
These values conform with the platykurtic range for the second
moment \cite{Stefanis:2015qha}
\begin{equation}
\langle \xi^2 \rangle_\pi^\text{pk}(\mu_2)
  = 0.220_{-0.006}^{+0.009}, ~~~
  \langle \xi^4 \rangle_\pi^\text{pk}(\mu_2)
  = 0.098_{-0.005}^{+0.008} \, ,
\label{eq:mom2-4-pk}
\end{equation}
%Eq (31)
while the fourth moment is also given for the sake of comparison with
other models.
One observes that the central point of the error contours is
not favored by the lattice simulations of \cite{Bali:2019dqc}.
(v) All positively correlated DAs are unimodal but have enhanced tails,
except the platykurtic one which shares tail suppression with the
anticorrelated BMS-like DAs.
(vi) Moreover, as we will see shortly, the platykurtic DA yields a TFF
in good agreement with all data compatible with strict scaling at large
$Q^2$ without crossing the pQCD asymptotic limit, cf.\ (\ref{eq:asy-TFF}).

Recently, some pion DAs have been proposed
\cite{Kaur_2020,qian2020light}
which yield moments
$\langle \xi^2 \rangle_\pi$, $\langle \xi^4 \rangle_\pi$
with values close to those of the platykurtic DA
in (\ref{eq:mom2-4-pk}) \cite{Stefanis:2015qha},
but employing different conceptions.

\subsection{TFF predictions}
\label{subsec:TFF-pred}
\begin{figure*}[t]
\includegraphics[width=0.48\textwidth]{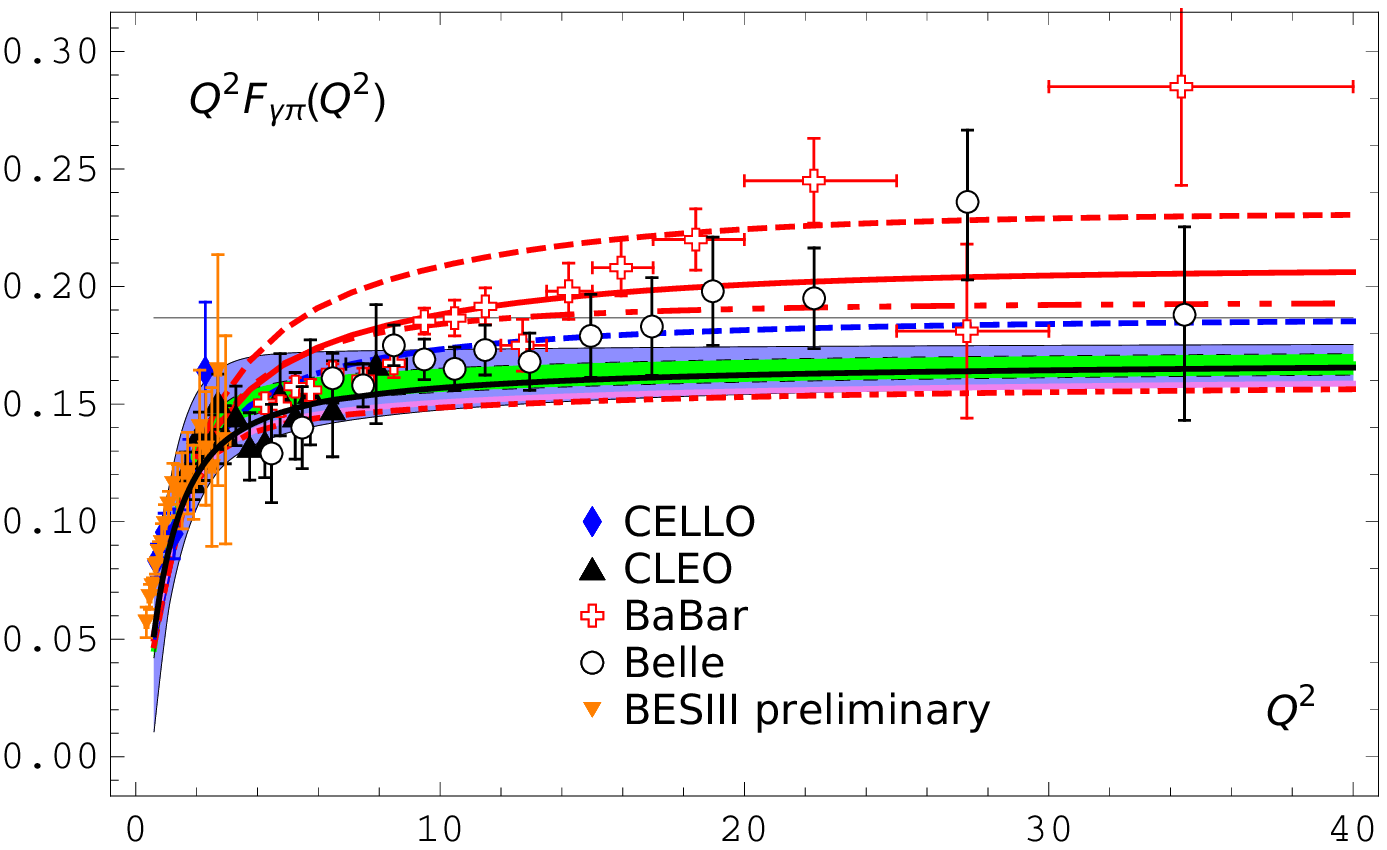}
\hfill
\includegraphics[width=0.48\textwidth]{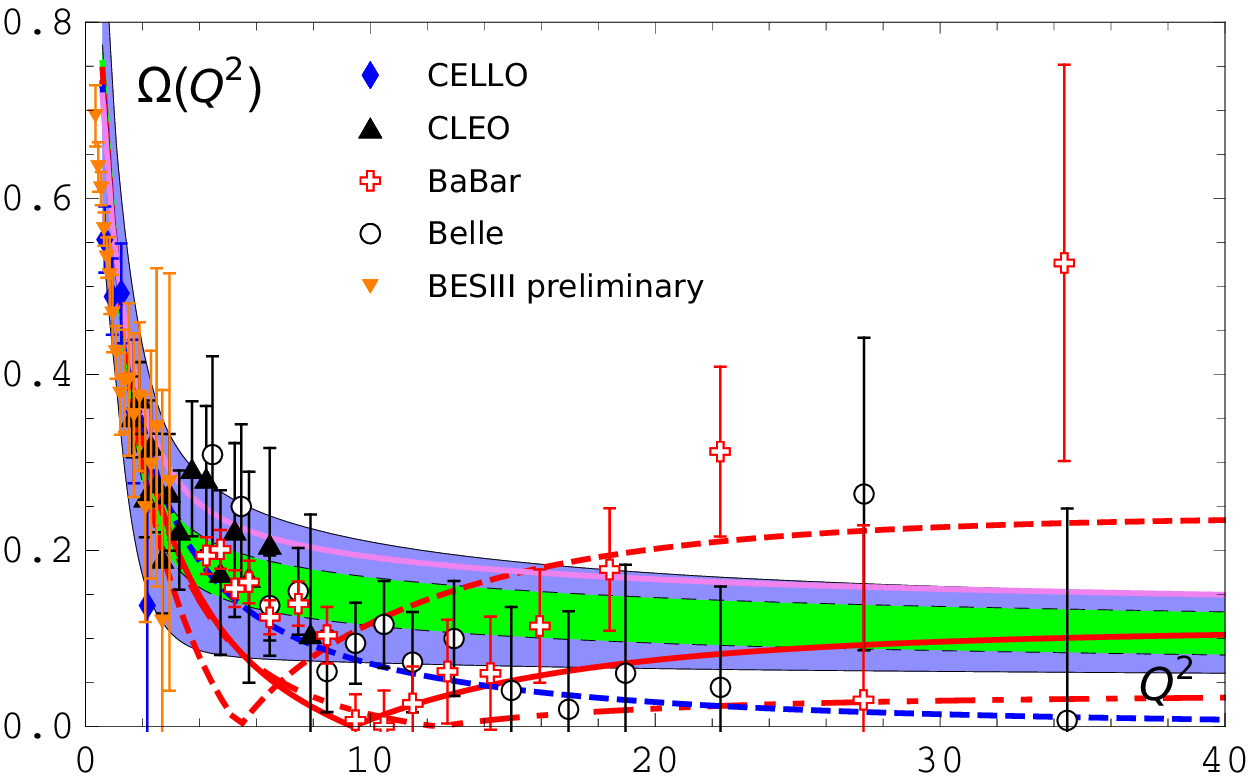}
\caption{Left: Measurements of the scaled pion-photon transition form
factor $Q^2F_{\gamma\pi}(Q^2)$ from different experiments in comparison
with theoretical predictions obtained from (\ref{eq:full-TFF})
according to Table \ref{tab:TFF-calculation} and using various pion
DAs defined in Table \ref{tab:pion-DAs-a6}.
The innermost (green) shaded strip shows the range of predictions
obtained with the bimodal BMS DAs from \cite{Bakulev:2001pa}.
The thick black line inside it denotes the result for the platykurtic
DA \cite{Stefanis:2014nla}.
The wider band (in blue color) around the green strip encapsulates the
principal theoretical uncertainty owing to the unknown NNLO term $T_c$,
see Eq.\ (\ref{eq:hard-scat-series}).
The upper two (red) lines illustrate the predictions from the DSE
approach:
DSE-DB \cite{Chang:2013pq} (solid line) and DSE-RL \cite{Chang:2013pq}
(dashed line).
The dashed-dotted-dotted (red) line denotes the prediction obtained
for the DSE-DB DA using a lower conformal resolution
(only $\{a_2,a_4\}$).
The dashed blue line below it represents the result derived from
AdS/QCD \cite{Brodsky:2011yv}, whereas the solid (pink) line and the
dashed-dotted (red) line below the lower boundary of the total BMS band
show the predictions calculated with a light-front quark model
\cite{Choi:2014ifm} and an instanton-based chiral quark model
\cite{Nam:2006sx}, respectively.
The horizontal solid line marks the asymptotic limit
$\mathcal{F}_{\infty} =\sqrt{2}f_\pi \approx 0.187$~GeV.
Right: Plot of the scaling-rate quantity $\Omega(Q^2)$,
cf.\ Eq.\ (\ref{eq:scaling-rate}),
in the $Q^2$ range $[0, 40]$~GeV$^2$ for various theoretical
TFF predictions in comparison with the data.
\label{fig:linear-scaled-TFF}
}
\end{figure*}

We now outline the calculational procedure to obtain predictions for
the scaled TFF $Q^2F_{\gamma\pi}(Q^2)$ using the pion DAs given in
Table \ref{tab:pion-DAs-a6}.
This discussion relies upon Table \ref{tab:TFF-calculation} in
correspondence with the formalism exposed in Sec.\ \ref{sec:theory}.
An extended discussion can be found in \cite{Stefanis:2019cfn}.
The results are shown in Fig.\ \ref{fig:linear-scaled-TFF} in comparison
with all existing data collected in App.\ \ref{sec:data-theory}.
The recently released preliminary data of the BESIII Collaboration
\cite{Redmer:2018uew,Ablikim:2019hff} (see also \cite{Danilkin:2019mhd})
are also included, keeping in mind that the probed momentum range extends
below 1~GeV$^2$, where our predictions are expected to be less reliable.

The calculated twist-two form factor is given in explicit form by
\begin{widetext}
\begin{eqnarray}
  F_\text{tw-2}^{\gamma^*\gamma^*\pi^0}(Q^2,q^2)
\! &\! = \! & \!
  N_\text{T} \Bigg[
              \underbrace{T_\text{LO}}_{(+)} + a_s(\mu^2)\underbrace{T_\text{NLO}}_{(-)}
            + a^2_s(\mu^2)\bigg(
                                  \underbrace{T_{{\rm NNLO}_{\beta_0}}}_{(-)}
                                + \underbrace{T_{{\rm NNLO}_{\Delta V}}}_{(-)}
                                + \underbrace{T_{{\rm NNLO}_{L}}}_{(0)}
                                + \underbrace{T_{{\rm NNLO}_{c}}}_{(?)}
                          \bigg)
                         + \ldots
      \Bigg] \!\otimes \varphi_\pi^{(2)}(x,\mu^2)
\nonumber \\
~~~~~~~~~~& + \! & \! \mathcal{O}\left(\frac{\delta^2}{Q^4}\right)
\label{eq:full-TFF}
\end{eqnarray}
\end{widetext}
%Eq (32)
with indications showing the sign of these contributions.
The label $(?)$ marks the only uncalculated NNLO term.

%%%%%%%%%%%%%%%%%%%%%%%%%%%%%%%%%%%%%%%%%%%%%%%%%%%%%%%%%%%%%%%%%%%%%%% Table II - Begin
\begin{center}
\begin{table*}[th]
\caption{Theoretical ingredients entering the TFF calculation within
the applied LCSR scheme using various pion DAs with conformal
coefficients $a_n$ ($a_0=1$) at
the normalization scales
$\mu_1=1$~GeV and
$\mu_2=2$~GeV given in Table \ref{tab:pion-DAs-a6}.
The question mark indicates that $\mathcal{T}_c $ is unknown.
It is included as the main theoretical uncertainty in the TFF
predictions obtained with the BMS/platykurtic DAs within FOPT,
see App.\ \ref{sec:data-theory}.
The other NNLO terms and the NLO contribution are explained in
Sec.\ (\ref{subsec:factorization}).
$\mathcal{F}_{\infty}$ denotes the asymptotic limit given by
Eq.\ (\ref{eq:asy-TFF}).
}
\smallskip
\smallskip
\begin{ruledtabular}
\begin{tabular}{lcccccc}
 LCSR $\left[\rho_2,\rho_4,\rho_6\right]$
        & LO+NLO
           & NNLO $(\alpha_{s}^{2})$
               & Error
                      & ERBL
                            & $\mathcal{F}_{\infty}$
                              &
\\ \phantom{.}
$\pi$ DAs & $T_\text{LO}+\alpha_{s}T_\text{NLO}$ & $T_\beta, ~~ T_{\Delta V}, ~~ T_L, ~~ \mathcal{T}_c $ & range & App.\ \ref{sec:global-NLO-evolution}
& Fig.\ \ref{fig:linear-scaled-TFF} (Left)&
\\
\hline               %MODEL                   %LO+NLO                   %NNLO                  %ERROR                   %ERBL/NLO global     %$\mathcal{F}_\infty$
 \vspace{-0.3cm}\\
 BMS \cite{Bakulev:2001pa} / pk \cite{Stefanis:2014nla} & $\left\{a_2,a_4\right\}_{\mu_1}$
 & ~~~~ $\left\{a_2,a_4\right\}_{\mu_1}$, $a_0, $ $0$, ?  & $\mathcal{T}_c \sim T_\beta$        & YES                      & below
 &   \\
 DSE \cite{Chang:2013pq,Raya:2015gva}$\left\{^\text{\small DB}_\text{\small RL}\right.$
 &  $\left\{a_2, a_4, \ldots, a_{12}\right\}_{\mu_2}$ &
 $\left\{a_2, a_4, a_6\right\}_{\mu_2}$, $a_0$, $0$, ? & NO                       & YES                    & above                                              &
  \\
 AdS/QCD \cite{Brodsky:2011yv} & $\left\{a_2, a_4, \ldots, a_{12}\right\}_{\mu_1}$& $\left\{a_2, a_4, a_6\right\}_{\mu_1}$, $a_0$, $0$, ? &   NO                 & YES
 & below                            &  \\
 Light-Front QM \cite{Choi:2014ifm} &  $\left\{a_2, a_4, a_6\right\}_{\mu_1}$ & $\left\{a_2, a_4, a_6\right\}_{\mu_1}$, $a_0$, 0, ?      &   NO                 & YES
 & below                            &  \\
 NL$\chi$ QM \cite{Nam:2006sx}      &  $\left\{a_2, a_4, a_6\right\}_{\mu_1}$ & $\left\{a_2, a_4, a_6\right\}_{\mu_1}$, $a_0$, 0, ?      &   NO                 & YES  & below                            &  \\
\end{tabular}
\end{ruledtabular}
\label{tab:TFF-calculation}
\end{table*}
\end{center}
%%%%%%%%%%%%%%%%%%%%%%%%%%%%%%%%%%%%%%%%%%%%%%%%%%%%%%%%%%%%%%%%%%%%%%% Table II - End

To facilitate the use of Table \ref{tab:TFF-calculation}, we briefly
describe the TFF calculation using as a reference model the set of the
BMS DAs determined in \cite{Bakulev:2001pa}.
These DAs are sufficiently well parameterized by means of the two
lowest coefficients $a_2$ and $a_4$, whereas higher coefficients
$a_{n>4}$, $(n=6, 8, 10)$ can be ignored because they were found to be
negligible albeit bearing large uncertainties \cite{Bakulev:2001pa}:
$a_6\approx a_2/3$;
$a_8\approx a_2/4$;
$a_{10}\approx a_2/5$.
The variation of $a_2, a_4$ allowed by the employed QCD sum rules gives
rise to the narrower (green) strip of predictions
in Fig.\ \ref{fig:linear-scaled-TFF}.
It includes at the twist-two level the LO, NLO, and NNLO-$T_{\beta_0}$
contributions to the short-distance coefficients, cf.\ (\ref{eq:T}),
(\ref{eq:hard-scat-nnlo-beta}), and involves the eigenfunctions
$\{\psi_0, \psi_2, \psi_4\}$.
The $\psi_0$ eigenfunction yields the largest (negative)
NNLO-$T_{\beta_0}$ contribution as we have seen in
Fig.\ \ref{fig:harmonics}.
Therefore, also the term $T_{\Delta V}\ll T_{\beta_0}$,
cf.\ (\ref{eq:hard-scat.nnlo-dv}),
is taken into account only via the zero harmonic $\psi_0$,
whereas the term NNLO-$T_{L}$ vanishes for $\psi_0$,
cf.\ (\ref{eq:hard-scat-nnlo}) \cite{Mikhailov:2016klg}.
The remaining NNLO term $T_c$ is unknown and this unknownness induces
the dominant theoretical uncertainty in the TFF prediction, shown in
terms of the broader (blue) band enveloping the narrower (green) one.
To gauge it, we assume that this term may be comparable in magnitude
to the leading NNLO term $T_{\beta_0}$
(likely overestimating its significance),
and obtain the uncertainties shown in the last column of
Table \ref{tab:ff-values-table}.
Table \ref{tab:ff-values-table} with some explanations on the data 
evaluation is given in Appendix \ref{sec:data-theory}.

Estimates of further theoretical errors---not considered here---can be
found in \cite{Stefanis:2012yw,Mikhailov:2016klg}.

The total TFF also comprises in the spectral density
(\ref{eq:rho-bar-15}) the contributions (\ref{eq:rho-tw-4})
(twist four) and (\ref{eq:tw-6}) (twist six) and includes NLO evolution
with heavy-quark crossings, see
App.\ \ref{sec:global-NLO-evolution}.
The calculation of the TFF with the platykurtic DA is similar
(black solid line in Fig.\ \ref{fig:linear-scaled-TFF}).
The analogous computations with the other considered pion DAs include
more conformal coefficients as indicated in
Table \ref{tab:TFF-calculation}.
The last row in this table provides information on the consistency of
each TFF prediction with the asymptotic limit
from pQCD \cite{Lepage:1980fj,Brodsky:1981rp},
\begin{equation}
  \lim_{Q^2\to\infty} \mathcal{F}(Q^2)
  =
   \sqrt{2} f_\pi
  \approx 0.187~\mbox{GeV} \, ,
\label{eq:asy-TFF}
\end{equation}
%Eq (33)
where we used the convenient notation
\begin{equation}
  Q^2F^{\gamma^*\gamma\pi^0(Q^2)}\equiv\mathcal{F}(Q^2) \, .
\label{eq:scaled-TFF}
\end{equation}
%Eq (34)

%%%%%%%%%%%%%%%%%%%%%%%%%%%%%%%%%%%%%%%%%%%%%%%%%%%%%%%%%%%%%%%%%%%%%%% Figure 4 - Begin
%%%%%%%%%%%%%%%%%%%%%%%%%%%%%%%%%%%%%%%%%%%%%%%%%%%%%%%%%%%%%%%%%%%%%%% Figure 4 - End

The calculated TFF predictions are collected in
Fig.\ \ref{fig:linear-scaled-TFF} (left panel) and are shown
in comparison with all data currently available:
CELLO \cite{Behrend:1990sr},
CLEO \cite{Gronberg:1997fj},
\textit{BABAR} \cite{Aubert:2009mc},
Belle \cite{Uehara:2012ag} 
(see Table \ref{tab:ff-values-table} in Appendix \ref{sec:data-theory}).
The BESIII data are given in \cite{Redmer:2018uew,Ablikim:2019hff}.
Apart from the results obtained with the BMS/pk DAs, already mentioned,
this figure contains the following curves.
The dashed red line farthest to the top shows $\mathcal{F}(Q^2)$ for
the DSE-RL DA \cite{Chang:2013pq}, whereas the solid red line below it
gives the result for the DSE-DB DA \cite{Chang:2013pq} using
$\{a_2, a_4, \ldots, a_{12}\}$.
To exhibit the influence of the attenuation effect on the TFF, we also
show the DSE-DB DA-based prediction using only $a_2$ and $a_4$
(long-dashed-dotted-dotted red line).
Obviously, this prediction agrees better with the asymptotic limit
(horizontal black line).
The reason is that it does not receive uninhibited contributions from
the higher harmonics with $a_{n>4}$ at higher $Q^2$, see
Fig.\ \ref{fig:harmonics}.
The computation of the TFF for the holographic DA \cite{Brodsky:2011yv}
yields the prediction represented by the dashed blue line running close
to the asymptotic limit.
Figure \ref{fig:linear-scaled-TFF} also includes the form-factor
predictions derived with the DA from the light-front quark model
\cite{Choi:2014ifm} (solid pink line) and the nonlocal chiral quark
(NL$\chi$QM) (model 3 in Table II \cite{Nam:2006sx})---dashed-dotted
red line.
Both lines run below all data above 8~GeV$^2$.

\subsection{TFF Asymptotics}
\label{subsec:TFF-asy}
The behavior of the form factor is known theoretically in two limits.
At $Q^2\to 0$ and in the chiral limit of quark masses, one obtains
from the axial anomaly \cite{Adler:1969gk,Bell:1969ts}
\begin{equation}
  \lim_{Q^2\to 0}F^{\gamma^*\gamma\pi^0}(Q^2)
=
  \frac{1}{2\sqrt{2}\pi^2 f_{\pi}} \, ,
\label{eq:axial-anomaly}
\end{equation}
%Eq (35)
where $f_{\pi}=132$~MeV is the leptonic decay constant of the pion.
On the other hand, the asymptotic behavior of the form factor is given
by Eq.\ (\ref{eq:asy-TFF}) which is an exact expression from pQCD.
The TFF in the $Q^2$ range between the aforementioned limits can be
phenomenologically described by the interpolation formula of Brodsky
and Lepage \cite{Brodsky:1981rp},
\begin{equation}
  F^{\gamma^*\gamma \pi}(Q^2)
=
  \frac{\sqrt{2}f_\pi}{4\pi^2f_\pi^2 + Q^2 } \ .
\label{eq:interpol}
\end{equation}
%Eq (36)

In order to study the scaling behavior of the calculated TFF at
large $Q^2$ more quantitatively, it is convenient to define the
quantity
\begin{equation}
  \Omega(Q^2)
\equiv
  \frac{\left|\mathcal{F}(Q^2) - \mathcal{F}_{\infty}\right|}
  {\mathcal{F}_{\infty}} \, ,
\label{eq:scaling-rate}
\end{equation}
%Eq (37)
which provides a normalized measure of the deviation of the scaled
form factor from the asymptotic value---the ``baseline''.
The graphical representation of the theoretical results for this
quantity versus $Q^2$ is shown in comparison with the data in
the right panel of Fig.\ \ref{fig:linear-scaled-TFF}.

Inspection of the graphics in Fig.\ \ref{fig:linear-scaled-TFF}
leads to the following observations: \\
(1) Most of the Belle data---except at $Q^2=27.33$~GeV$^2$---come
within errors close to the baseline clearly indicating that the TFF
approaches an asymptotic value.
Indeed, using the dipole formula
$\mathcal{F}(Q^2)=BQ^2/(C+Q^2)$
the Belle Collaboration determined $B=0.209 \pm 0.016$~GeV, which is
slightly larger than the exact result $\sqrt{2}f_\pi$, but still
compatible. \\
(2) Several \textit{BABAR} data points above 10~GeV$^2$ do not
indicate TFF saturation as they move away from $\mathcal{F}_\infty$.
Because also their error bars do not reach the $\mathcal{F}_\infty$
baseline, one may think that this data deviation is systematic and
self-generated.
It was shown in \cite{Stefanis:2012yw} that a dipole best-fit to the
\textit{BABAR} data yields $B=0.23$~GeV and $C=2.6$~GeV$^2$ with
$\chi^2=1.7$, being unable to reproduce the Belle data with acceptable
accuracy.
The origin of these discrepancies is the subject of several theoretical
investigations (see \cite{Bakulev:2012nh} for a detailed discussion and
references). \\
(3) The $Q^2$ intervals above 20~GeV$^2$, probed by \textit{BABAR}
and Belle, are only scarcely populated and have a rather poor
statistics. \\
(4) The recently released preliminary BESIII data
\cite{Redmer:2018uew,Ablikim:2019hff} cover the momentum range
$[0.3-3.1]$~GeV$^2$ and exceed the statistical accuracy of the CELLO
data at $Q^2\lesssim 1$~GeV$^2$ considerably, though their error bars
become larger in the range $1.5\leqslant 3.1$~GeV$^2$.
These results are important for the hadronic light-by-light scattering
calculations \cite{Hoferichter:2018dmo,Hoferichter:2018kwz,%
Masjuan:2017tvw,Guevara:2018rhj,Aoyama:2020ynm}.
As regards the TFF below 1~GeV$^2$, other dispersion approaches may be
more adequate
\cite{Klopot:2012hd,Oganesian:2015ucv,Ayala:2018ifo,Ayala:2019etj}
than the LCSR scheme applied in this work. \\
(5) The broader (blue) band of predictions obtained with the BMS DAs
(including their main uncertainties)
approaches $\mathcal{F}_\infty$ gradually from lower values without
reaching it (similarly also the platykurtic prediction).
In the momentum interval $[10-11]$~GeV$^2$, the BMS TFF takes the
value
$\mathcal{F}_\text{BMS}(Q^2)=0.1604_{+0.0128}^{-0.124}$~GeV,
while the pk TFF reaches this value
at $[13.5 - 15.0]$~GeV$^2$
(see Table \ref{tab:ff-values-table} in Appendix \ref{sec:data-theory}).
Both TFFs grow very slowly towards higher $Q^2$ values indicating
that they have already entered the pre-asymptotic regime due to
saturation. \\
(6) The TFF predictions associated with the LFQM-based DA
\cite{Choi:2014ifm} (solid line) and the NL$\chi$ QM DA
\cite{Nam:2006sx} (dashed-dotted red line), respectively, follow
the trend of the platykurtic-generated prediction but have smaller
magnitudes so that the onset of pre-asymptotic behavior in the TFF
is shifted to much higher momenta.
This is because both models have a negative $a_6$ coefficient
(see Table \ref{tab:pion-DAs-a6}) that slows the TFF saturation. \\
(7) The prediction based on the holographic AdS/QCD DA
\cite{Brodsky:2011yv} provides a rather good agreement with the Belle
data, while it disagrees with the \textit{BABAR} data above
10~GeV$^2$, where these start to grow.
However, it crosses the baseline at still higher $Q^2$
around 57~GeV$^2$.
Variants of the holographic DA in \cite{Chang:2016ouf} either
yield similar results or tend to cross the baseline quite fast. \\
(8) Still broader concave pion DAs, like DSE-DB and DSE-RL,
lead to predictions that reach the pre-asymptotic regime in a way
sensitive to the power $\alpha_{-}$ in the ``Gegenbauer-$\alpha$''
representation.
This employs Gegenbauer polynomials of variable dimensionality
$\alpha=\alpha_{-}+1/2$
\cite{Chang:2013pq,Gao:2014bca}
\begin{equation}
  \varphi_{\pi}^{(\alpha)}(x,\mu^2)
=
  N_\alpha (x\bar{x})^{\alpha_{-}}
  [1 + a_{2}^{\alpha}C_{2}^{(\alpha)}(x-\bar{x})]
\label{eq:DSE-DA}
\end{equation}
%Eq (38)
and gives TFF predictions with magnitudes growing in inverse proportion
to $\alpha_{-}$.
Employing the set $\{a_2, a_4, \ldots , a_{12}\}$, both
TFF predictions cross the $\mathcal{F}_\infty$ line already at
$Q^2\approx 4$~GeV$^2$ (DSE-RL) and
$Q^2\approx 10$~GeV$^2$ (DSE-DB)
and continue to grow.
The reduced DSE-DB, which uses only $a_2,a_4$, leads to a TFF with a
better asymptotic behavior (dashed-dotted-dotted red line) entering
the pre-asymptotic regime around 10~GeV$^2$ and remaining then close
to $\mathcal{F}_\infty$ but above it. \\
(9) The theoretical predictions depend crucially on the DA models
involved in the calculation.
An attempt to extract the asymptotic behavior of the TFF directly
from the data was given in \cite{Stefanis:2019cfn}.

\section{Summary and Outlook}
\label{sec:summary}
In this work we carried out a comprehensive analysis of the pion-photon
transition form factor in QCD using the method of LCSRs within FOPT to
NNLO and twist-six accuracy.
The presented predictions for this exclusive observable are of
considerable interest for two different reasons:
(i) they provide a handle on the involved pion distribution amplitude
and (ii) they represent a powerful tool to study the onset of scaling
at high $Q^2$ in present-days experiments.

To analyze in detail the $Q^2$ behavior of the TFF within the LCSR
approach in comparison to pQCD, we studied the attenuation effect of
the partial components of the TFF related to the conformal
expansion of the pion DA and worked out how they contribute as $Q^2$
grows.
We showed that broad, concave DAs relying on many positive conformal
coefficients will tend to exceed the asymptotic limit
$\mathcal{F}_\infty$
because higher components $F_n(Q^2)$ will start to contribute far
beyond 10~GeV$^2$.

A good agreement with $\mathcal{F}_\infty$ presumes saturation of the
scaled TFF at large $Q^2$ and entails the onset of scaling.
The big unknown is at which momentum scale this becomes obvious
\cite{Eichmann:2017wil,Hoferichter:2020lap}.
The over-all agreement of the TFF predictions, obtained in this work
with the set of the BMS DAs \cite{Bakulev:2001pa} (including the
platykurtic one \cite{Stefanis:2014nla}), with the
CLEO \cite{Gronberg:1997fj},
\textit{BABAR}($<9~\mbox{GeV}^2$) \cite{Aubert:2009mc}, and
Belle \cite{Uehara:2012ag}
data is good and any discrepancies are within the corresponding
experimental errors,
see Fig.\ \ref{fig:linear-scaled-TFF}.
However, they disagree with the \text{BABAR} data above 10~GeV$^2$.
These predictions are shown as shaded bands and include the principal
theoretical uncertainties of the conformal coefficients (narrower band)
and the incomplete knowledge of the NNLO radiative corrections
(wider band) in these figures.
Saturation is observed in the interval $[10-14]$~GeV$^2$, where
the TFF reaches the value $\gtrsim 0.16$~GeV, indicating the onset of
the pre-asymptotic regime.

More ambitiously, the combination of the $1\sigma$ and $2\sigma$ error
regions of all data compatible with asymptotic scaling with the most
recent lattice constraints from \cite{Bali:2019dqc} at the NNLO and NLO
level supports a platykurtic pion DA with
$a_2 \approx 0.08, a_4 \approx -0.009$
\cite{Stefanis:2014nla,Stefanis:2015qha}, though the sign and magnitude
of $a_4$ require further consideration in the lattice context or
otherwise.
From the experimental side, it would be very helpful to have more data
in regular steps of 1~GeV$^2$ above 10~GeV$^2$.
This would enable a reliable data-driven analysis based on the state-space
reconstruction method, proposed in \cite{Stefanis:2019cfn}, and we hope
that the Belle-II Collaboration will perform such measurements.

\acknowledgments
I would like to thank Sergey Mikhailov and Alex Pimikov for
a fruitful collaboration during the last years.
I am indebted to Alex Pimikov for help with the numerical computations
and their graphical representation.
I am grateful to Gunnar Bali for useful remarks.

\begin{appendix}
\section{NLO evolution of the pion DA including heavy-quark thresholds}
\label{sec:global-NLO-evolution}
In this Appendix
(done partly in collaboration with S.\ V.\ Mikhailov and A.\ V.\ Pimikov)
we discuss the NLO (i.e., the two-loop) ERBL
evolution of the pion DA with an arbitrary number of Gegenbauer
coefficients taking into account heavy-quark
flavors (also known as global QCD scheme, see, e.g.,
\cite{Bakulev:2012sm} and references therein).
This scheme employs the global coupling
$\alpha_{s}^\text{glob}(Q^2,\Lambda_{N_f}^2)$
that depends on the number of flavors $N_f$ through the QCD
scale parameter $\Lambda_{N_f}$.

This procedure was used in this work to derive the results given
in Tables \ref{tab:pion-DAs-a6}, \ref{tab:TFF-calculation}, and
\ref{tab:ff-values-table}
and obtain the predictions shown in all figures.
It takes into account the heavy-quark mass thresholds and thus requires
the matching of the strong coupling in the Euclidean region of $Q^2$ at
the corresponding heavy-quark masses when one goes from
$N_f \to N_{f}+1$.
Note that the dependence on $N_f$ in Appendix D of
\cite{Bakulev:2002uc},
which provided the basis for the NLO evolution of the pion DA in our
earlier works, was ignored assuming a fixed number of flavors.
The new scheme has already been used in our more recent investigations
\cite{Bakulev:2012nh,Stefanis:2012yw,Stefanis:2014yha,Mikhailov:2016klg},
but without exposing the underlying formalism in final form.
This task will be accomplished here including further refinements.
The NLO evolution of the pion DAs with two conformal coefficients
$a_2$ and $a_4$ at the initial scale $\mu^2\simeq 1$~GeV$^2$
and a varying number of heavy flavors has also been applied in
\cite{Bakulev:2004cu} (see Appendix D there).
Our technical exposition below extends this treatment to any number
of conformal coefficients and more heavy-flavor thresholds.
For some specific details and references, we refer to
\cite{Bakulev:2008td,Bakulev:2012sm}.

Let us start with a fixed number of flavors and supply some basic
formulas from \cite{Bakulev:2002uc}.
The ERBL evolution equation for the pion DA is given by
\begin{equation}
  \frac{d\,\varphi_{\pi}(x;\mu^2)}{d\,\ln\mu^2}
=
  V\left(x,u;a_s(\mu^2)\right)\underset{u}\otimes\varphi_{\pi}(u;\mu^2)
\label{eq:ERBL-eq}
\end{equation}
%Eq (A1)
and is driven by the kernel
\begin{equation}
  V(x,y;a_s)
=
  a_s~ V_0(x,y)
  + a_s^2~ V_1(x,y)
  + \ldots
\label{eq:kernel}
\end{equation}
%Eq (A2)
with $a_s=\alpha_s/(4\pi)$.

The eigenvalues $\gamma_{n}(a_s)$
and the one-loop eigenfunctions $\psi_{n}(u)$
are related to the kernel $V$ through
\begin{equation}
  \tilde{\psi}_{n}(x) \underset{x}\otimes V(x,u;a_s)\underset{u}\otimes \psi_{n}(u)
=
  - \gamma_n(a_s)\, ,
\label{eq:eigen}
\end{equation}
%Eq (A3)
where
$
 \displaystyle \tilde{\psi}_{n}(x)
=
 2(2n+3)/\left[3(n+1)(n+2)\right] C^{3/2}_n(x-\bar{x})
$.
The explicit expressions for the anomalous dimensions $\gamma_{n}$
at one loop, $\gamma_0(n)$,
and at two-loops, $\gamma_1(n)$,
in the expansion
$
 \gamma_{n}(a_s)
=
 \frac{1}{2}[a_s\gamma_{0}(n) + a_{s}^2\gamma_{1}(n) + \ldots]
$
can be found in Appendix D in \cite{Bakulev:2002uc}.

To perform the pion DA evolution, while ignoring quark-mass thresholds,
we make use of the evolution matrix $E$ with the components
$E_{nk}$.
Expanded over the basis $\{\psi_n\}$ of the Gegenbauer harmonics, this
matrix assumes the following triangular form \cite{Mikhailov:1985cm}
\begin{widetext}
\begin{eqnarray}
\label{eq:DA_EVO-glob}
  E_{nk}(N_f;Q^2,\mu^2)
&=&
  P(n,Q^2,\mu^2)\left[\delta_{nk}+ a_s(Q^2)
  \Theta(k-n>0)d_{nk}(Q^2,\mu^2)\right]\, , \\
d_{nk}(\mu^2,\mu^2) = 0 \, ,
\label{eq:e-norm}
\end{eqnarray}
%Eq (A4), (A5)
where the coefficients
$d_{nk}(Q^2,\mu^2)$
will be defined shortly, and where $\mu^2$ and $Q^2$ refer to
the initial and observation scale, respectively.
The factor $P(n,Q^2,\mu^2)$ in Eq.\ (\ref{eq:DA_EVO-glob}) denotes the
diagonal part of the evolution matrix that dominates the
renormalization-group (RG) controlled evolution of the
$\psi_n$-harmonics in the conformal expansion
\begin{eqnarray}
  \varphi_\pi^\text{RG}(x,Q^2)
&=& \sum\limits_{n} a_{n}(\mu^2)\left\{ P(n,Q^2,\mu^2)
  \left[\psi_{n}(x) + a_{s}(Q^2)
  \sum\limits_{k>n} d_{nk}(Q^2,\mu^2) \psi_{k}(x)
  \right]\right\}\, .
\label{eq:DA_EVO}
\end{eqnarray}
%Eq (A6)
Then, the diagonal part of the evolution exponential at the
two-loop level can be given explicitly,
\begin{eqnarray}
  P(n,Q^2,\mu^2)
&=&
  \exp
      \left[
            \int\limits_{a_{s}(\mu^2)}^{a_{s}(Q^2)}
            \frac{\gamma_n(a)}{\beta(a)}da
      \right]
\stackrel{2-\text{loops}}{\longrightarrow}
      \left[
            \frac{a_s(Q^2)}{a_s(\mu^2)}
      \right]^{\frac{\gamma_0(n)}{2b_0}}
      \left[\frac{1+c_1a_s(Q^2)}{1+c_1a_s(\mu^2)}
      \right]^{\omega(n)}\, ,
\label{eq:Ediag}
\end{eqnarray}
%Eq (A7)
\end{widetext}
where
$a_s(\mu^2)= \alpha_s^{\text{glob};(2)}(\mu^2,\Lambda_3^2)/(4\pi)$
and  $c_1=b_1/b_0$, with $b_i$ being the expansion coefficients of the
QCD $\beta$-function.
The evolution exponent of the coupling is defined by
$
 \omega(n)
=
 [\gamma_1(n)b_0-\gamma_0(n)b_1]/[2b_0b_1]
$.
The second term in the brackets in Eq.\ (\ref{eq:DA_EVO-glob})
represents the non-diagonal part of the evolution equation to the
order $O(a_s^2)$ induced by renormalization and encodes the mixing
of the higher Gegenbauer harmonics for indices $k > n$ related to
the conformal-symmetry breaking at NLO \cite{Mueller:1994cn}.
Notice that all components on the right-hand side of
Eqs.\ (\ref{eq:DA_EVO-glob}) and (\ref{eq:Ediag}) depend on $N_f$,
which changes to $N_f+1$, when the next quark-mass threshold is
crossed.
The explicit form of the mixing coefficients is given by
\cite{Bakulev:2002uc}
\begin{widetext}
\begin{eqnarray}
 d_{nk}(Q^2,\mu^2)
  & = &
  \frac{M_{nk}}{\gamma_0(k)-\gamma_0(n)-2b_0}
  \left\{1
    - \left[\frac{a_s(Q^2)}
                 {a_s(\mu^2)}
      \right]^{[\gamma_0(k)-\gamma_0(n)]/(2b_0)-1}
  \right\}\, ,
\end{eqnarray}
%Eq (A8)
where the values of the first few elements
of the matrix $M_{nk}$ ($k=2, 4\geq n=0, 2$) read
\begin{eqnarray}
\label{eq:App-Mnk}
  M_{0 2} = -11.2 + 1.73  N_f, ~~
  M_{0 4} = -1.41 + 0.565 N_f, ~~
  M_{2 4} = -22.0 + 1.65  N_f\, .
\end{eqnarray}
%Eq (A9)
\end{widetext}
Analytic expressions for $M_{nk}$ have been obtained
in \cite{Mueller:1993hg}.
The values in Eq.\ (\ref{eq:App-Mnk}) reproduce the exact
results with a deviation less than about $1\%$.

To make our further exposition more compact, we make use of the
parameter vectors
$\mathbf{A}_{}(\mu^2)$ and $\bm{\Psi}(x)$
defined at the reference momentum scale $\mu^2$ as follows
\begin{subequations}
\begin{eqnarray}
  \mathbf{A}
&=&
  (1,\,a_2,\,a_4\,,\cdots,\,a_{2(N-1)})\, ,
\\
  \bm{\Psi}
&=&
  (\psi_0,\psi_2,\cdots,\psi_{2(N-1)})\, ,
\\
  \varphi_{\pi}(x,\mu^2)
&=&
  \sum\limits_{n=0}^{N-1} a_{2n}(\mu^2)\psi_{2n}(x)
\nonumber \\
&=&
  \mathbf{A}_{}(\mu^2)\bm{\Psi}(x)\,,
\end{eqnarray}
\end{subequations}
%Eq (A10a), (A10b), (A10c)
where their dimension and the dimension of the matrix $E$ depends
on the parameter $N$.
Then, the evolution of the pion DA can be carried out in terms of the
Gegenbauer coefficients $a_i$ with $i=2, 4,\ldots ,2(N-1)$.
For a fixed number of flavors, one gets
\begin{subequations}
\begin{eqnarray}
  \bm{\Psi}(x;\mu^2)
& = &
  E(N_f,\mu^2,\mu^2_0)\bm{\Psi}(x), \\
\mathbf{A}(\mu^2)
& = &
  E^\text{T}(N_f,\mu^2,\mu^2_0)\mathbf{A}(\mu_0^2)\, ,
\end{eqnarray}
\end{subequations}
%Eq (A11a), (A11b)
where $E^\text{T}$ is the transposed matrix of $E$, while
$\mathbf{A}(\mu_0^2)$ is the vector of the Gegenbauer coefficients
defined at some initial scale $\mu_0^2$.

In the global QCD scheme, the evolution of the pion DA defined at the
initial scale $\mu_0^2$, is implemented by means of the threshold
interval factors $E_i$ in the following step-by-step procedure,
\begin{widetext}
\begin{eqnarray}
  E_\text{glob}(\mu^2,\mu^2_0)
&=&
   E_3(\mu^2)\theta(\mu^2<M_4^2)
  +
   E_4(\mu^2)\theta(M_4^2\leqslant\mu^2<M_5^2)E_3
  + \nonumber \\
&&
   E_5(\mu^2)\theta(M_5^2\leqslant\mu^2<M_6^2)E_4E_3
  +
   E_6(\mu^2)\theta(M_6^2\leqslant\mu^2)E_5E_4E_3 \, ,
   \label{eq:global-pionDA-evo}
\end{eqnarray}
%Eq (A12)
where the matrices $E_i$ and $E_i(\mu^2)$ are given by
\begin{eqnarray}
  E_i(\mu^2)
\equiv
  E(i,\mu^2,M^2_i) \, ,~~
  E_i
\equiv
  E(i,M^2_{i+1},M^2_i)
\end{eqnarray}
%Eq (A13)
\end{widetext}
and the thresholds are defined \cite{Bakulev:2012sm} by the heavy-quark
masses
$m_c\sim M_4=1.65$~GeV,
$m_b \sim M_5=4.75$~GeV,
and $m_t\sim M_6 = 172.5$~GeV,
while
$M^2_3\equiv \mu^2_0$ sets the initial scale
taken to be either $\mu_{0}=\mu_{1}=1$~GeV or $\mu_{0}=\mu_2=2$~GeV,
see Table \ref{tab:pion-DAs-a6}.
Note that the global evolution matrix,
Eq.\ (\ref{eq:global-pionDA-evo}),
is presented for $\mu_0<M_4$ and $\mu>\mu_0$.
No \textit{matching} at the mass thresholds is needed in the case of
equal initial and final momentum scales, i.e., $E(N_f;Q^2,Q^2)=1$
because of the independence of the evolution matrix on the number
of flavors.
For example, at the threshold $M_4$, we have $E_4(M_4^2)=1$
ensuring the continuity of the global evolution matrix $E_\text{glob}$.
It is worth noting that our NLO evolution scheme in terms
of Eq.\ (\ref{eq:global-pionDA-evo}),
has the following improvements relative to that used in
\cite{Bakulev:2004cu}
(see Appendix D there):
(a)	It is applicable to DAs with any number of Gegenbauer harmonics.
(b) The number of heavy-quark thresholds is extended to four flavors.
(c) When the interval of evolution contains two or more mass thresholds,
    our method can still incorporate contributions from the
    non-diagonal part of the evolution matrix, removing the restriction
    of using only the first two Gegenbauer coefficients $a_2$ and $a_4$
    as in \cite{Bakulev:2004cu}.

We reiterate that the matching of the coupling constants at the
quark-mass thresholds requires the readjustment of the value of the
QCD scale parameter
$\Lambda$ to $\Lambda_{(N_f)}$.
A detailed description of the matching procedure of the
running coupling in the global scheme can be found, for instance, in
\cite{Bakulev:2012sm}.
For definiteness, we quote here the two-loop $\Lambda_{(N_f)}^{(2)}$
values used in our code:
$\Lambda_{(3)}^{(2)} = 369$~MeV,
$\Lambda_{(4)}^{(2)} = 305$~MeV,
$\Lambda_{(5)}^{(2)} = 211$~MeV,
$\Lambda_{(6)}^{(2)} = 88$~MeV.
These values are defined by fixing the strong coupling
\begin{equation}
  \alpha_S(M_Z^2)=0.118
\end{equation}
%Eq (A14)
at the scale of the Z boson mass $M_Z=91$~GeV.

We emphasize that for self-consistency reasons, the
global two-loop coupling
$\alpha_s^{\text{glob};(2)}(\mu^2,\Lambda_3)/(4\pi)$
should be used in \textit{all} functions entering
Eq.\ (\ref{eq:DA_EVO-glob}) that depend on the coupling with a variable
flavor number $N_f$.
Finally, the global evolution of the Gegenbauer coefficients is given
by
\begin{eqnarray}
  \mathbf{A}_\text{glob}(\mu^2)
=
  E^\text{T}_\text{glob}(\mu^2,\mu^2_0)\mathbf{A}(\mu_0^2)\, ,
\end{eqnarray}
%Eq (A14)
whereas the global evolution of the pion DA assumes the form
\begin{widetext}
\begin{eqnarray}
  \varphi_{\pi}^\text{glob}(x,\mu^2)
&=&
  \mathbf{A}_{\text{glob}}(\mu^2)\bm{\Psi}(x)= \bm{A}(\mu_0^2)\bm{\Psi}(x;\mu^2)
=
  \sum\limits_{n=0}^{N-1} a_{2n}^{\text{glob}}(\mu^2)\psi_{2n}(x)
\, .
\label{eq:DA_EVO-glob-final}
\end{eqnarray}
%Eq (A15)
\end{widetext}

\section{Experimental data and numerical predictions}
\label{sec:data-theory}
In this appendix, we collect the experimental data on the
pion-photon TFF together with our main theoretical predictions.
The preliminary BESIII data \cite{Redmer:2018uew,Ablikim:2019hff}
are included in the graphics shown in Fig.\ \ref{fig:linear-scaled-TFF}.
The results obtained with the BMS pion DA \cite{Bakulev:2001pa}
in the last column differ from those we reported before in
\cite{Bakulev:2011rp} because here we used the updated
theoretical framework discussed in Sec.\ \ref{sec:theory} and
App.\ \ref{sec:global-NLO-evolution}.
The numbers given in parentheses are new predictions calculated with
the platykurtic pion DA \cite{Stefanis:2014nla}.
Their theoretical uncertainties are covered by those for the BMS DAs.
%%%%%%%%%%%%%%%%%%%%%%%%%%%%%%%%%%%%%%%%%%%%%%%%%%%%%%%%%%%%%%%%%%%%%%% TABLE III - Begin
\begin{table*}[]
\begin{center}
\begin{ruledtabular}
\caption{Compilation of the existing data on
$
 \tilde{Q}^2F^{\gamma^{*}\gamma\pi^0}(\tilde{Q}^2)
\equiv
 \mathcal{F}_{\gamma\pi}(\tilde{Q}^2)
$
from single-tag experiments:
CELLO \protect\cite{Behrend:1990sr},
CLEO \protect\cite{Gronberg:1997fj},
\textit{BABAR} \protect\cite{Aubert:2009mc},
and Belle \cite{Uehara:2012ag}.
The TFF is measured at $\tilde{Q}^2$ where the differential cross
sections assume their mean values computed by numerical integration.
The last column shows theoretical predictions and uncertainties for
the BMS and pk DAs.
\label{tab:ff-values-table}}
\smallskip
\smallskip
\begin{tabular}{ccccccc}
%Exp & Exp & CELLO & CLEO  & \babar & Belle & BMS DA \\
 $Q^2$ {\small bin range}
        & $\tilde{Q}^2$
           & $\mathcal{F}_\text{CELLO}^{\gamma^*\gamma\pi^0}(\tilde{Q}^2)$
               & $\mathcal{F}_\text{CLEO}^{\gamma^*\gamma\pi^0}(\tilde{Q}^2)$
                      & $\mathcal{F}_\text{BABAR}^{\gamma^*\gamma\pi^0}(\tilde{Q}^2)$
                            & $\mathcal{F}_\text{Belle}^{\gamma^*\gamma\pi^0}(\tilde{Q}^2)$
                                   & $\mathcal{F}_{\rm BMS(pk)}^{\gamma^*\gamma\pi^0}(\tilde{Q}^2)$
\\ \phantom{.}
[GeV$^2$] & [GeV$^2$] & [0.01 $\times$ GeV] & [0.01 $\times$ GeV] & [0.01 $\times$ GeV] & [0.01 $\times$ GeV] & [0.01 $\times$ GeV]
\\
\hline               %CELLO                    %CLEO                      %BABAR                  %Belle                           %BMS(pk)
 0.5 --  0.8 &  0.68 &  8.37$_{-0.73}^{+0.67}$ & --                       & --                    & --                             &  5.39$_{-3.37}^{+3.46}$(5.98)  \\
 0.8 --  1.1 &  0.94 &  9.58$_{-0.84}^{+0.78}$ & --                       & --                    & --                             &  7.70$_{-2.80}^{+2.90}$(7.95)  \\
 1.1 --  1.5 &  1.26 &  9.54$_{-1.12}^{+1.00}$ & --                       & --                    & --                             &  9.94$_{-2.50}^{+2.62}$(9.65)  \\
 1.5 --  1.8 &  1.64 & --                      & 12.1$\pm\,0.8\,\pm\,$0.3 & --                    & --                             & 11.78$_{-2.46}^{+2.6} $(11.05) \\
 1.5 --  2.1 &  1.70 & 12.08$_{-1.62}^{+1.43}$ & --                       & --                    & --                             & 12.00$_{-2.45}^{+2.59}$(11.23) \\
 1.8 --  2.0 &  1.90 & --                      & 11.7$\pm\,0.7\,\pm\,$0.3 & --                    & --                             & 12.66$_{-2.39}^{+2.55}$(11.75) \\
 2.0 --  2.2 &  2.10 & --                      & 13.8$\pm\,0.8\,\pm\,$0.3 & --                    & --                             & 13.18$_{-2.32}^{+2.49}$(12.19) \\
 2.1 --  2.7 &  2.17 & 16.43$_{-3.60}^{+2.94}$ & --                       & --                    & --                             & 13.33$_{-2.29}^{+2.46}$(12.33) \\
 2.2 --  2.4 &  2.30 & --                      & 12.7$\pm\,0.9\,\pm\,$0.3 & --                    & --                             & 13.59$_{-2.24}^{+2.42}$(12.56) \\
 2.4 --  2.6 &  2.50 & --                      & 13.5$\pm\,1.0\,\pm\,$0.3 & --                    & --                             & 13.93$_{-2.16}^{+2.34}$(12.88) \\
 2.6 --  2.8 &  2.70 & --                      & 15.1$\pm\,1.1\,\pm\,$0.4 & --                    & --                             & 14.20$_{-2.07}^{+2.25}$(13.16) \\
 2.8 --  3.1 &  2.94 & --                      & 13.7$\pm\,1.2\,\pm\,$0.3 & --                    & --                             & 14.46$_{-2.01}^{+2.18}$(13.43) \\
 3.1 --  3.5 &  3.29 & --                      & 14.5$\pm\,1.2\,\pm\,$0.4 & --                    & --                             & 14.75$_{-1.93}^{+2.09}$(13.76) \\
 3.5 --  4.0 &  3.74 & --                      & 13.2$\pm\,1.4\,\pm\,$0.3 & --                    & --                             & 15.01$_{-1.84}^{+1.99}$(14.10) \\
 4.0 --  4.5 &  4.24 & --                      & 13.4$\pm\,1.5\,\pm\,$0.3 & 15.04$\pm\,0.39$      & --                             & 15.21$_{-1.76}^{+1.88}$(14.39) \\
 4.0 --  5.0 &  4.46 & --                      & --                       & --                    & 12.9$\pm\,2.0\,\pm\,$0.6       & 15.28$_{-1.72}^{+1.84}$(14.50) \\
 4.5 --  5.0 &  4.74 & --                      & 15.4$\pm\,1.7\,\pm\,$0.4 & 14.91$\pm\,0.41$      & --                             & 15.36$_{-1.68}^{+1.79}$(14.62) \\
 5.0 --  5.5 &  5.24 & --                      & 14.5$\pm\,1.8\,\pm\,$0.4 & 15.74$\pm\,0.39$      & --                             & 15.48$_{-1.61}^{+1.71}$(14.81) \\
 5.0 --  6.0 &  5.47 & --                      & --                       & --                    & 14.0$\pm\,1.6\,\pm\,$0.7       & 15.52$_{-1.58}^{+1.68}$(14.89) \\
 5.5 --  6.0 &  5.74 & --                      & 15.5$\pm\,2.2\,\pm\,$0.4 & 15.60$\pm\,0.45$      & --                             & 15.57$_{-1.55}^{+1.64}$(14.97) \\
 6.0 --  7.0 &  6.47 & --                      & 14.8$\pm\,2.0\,\pm\,$0.4 & 16.35$\pm\,0.36$      & 16.1$\pm\,0.7\,\pm\,$0.8       & 15.68$_{-1.48}^{+1.56}$(15.15) \\
 7.0 --  8.0 &  7.47 & --                      & --                       & 16.06$\pm\,0.47$      & 15.8$\pm\,0.6\,\pm\,$0.7       & 15.80$_{-1.40}^{+1.47}$(15.35) \\
 7.0 --  9.0 &  7.90 & --                      & 16.7$\pm\,2.5\,\pm\,$0.4 & --                    & --                             & 15.84$_{-1.37}^{+1.44}$(15.42) \\
 8.0 --  9.0 &  8.48 & --                      & --                       & 16.73$\pm\,0.60$      & 17.5$\pm\,0.5\,\pm\,$0.7       & 15.89$_{-1.34}^{+1.39}$(15.51) \\
 9.0 -- 10.0 &  9.48 & --                      & --                       & 18.53$\pm\,0.55$      & 16.9$\pm\,0.5\,\pm\,$0.7       & 15.97$_{-1.28}^{+1.33}$(15.63) \\
10.0 -- 11.0 & 10.48 & --                      & --                       & 18.66$\pm\,0.76$      & 16.5$\pm\,0.6\,\pm\,$0.7       & 16.04$_{-1.24}^{+1.28}$(15.73) \\
11.0 -- 12.0 & 11.48 & --                      & --                       & --                    & 17.3$\pm\,0.8\,\pm\,$0.7       & 16.10$_{-1.20}^{+1.24}$(15.82) \\
11.0 -- 12.0 & 11.49 & --                      & --                       & 19.16$\pm\,0.78$      & --                             & 16.10$_{-1.20}^{+1.24}$(15.82) \\
12.0 -- 13.5 & 12.71 & --                      & --                       & 17.50$\pm\,1.10$      & --                             & 16.16$_{-1.16}^{+1.19}$(15.91) \\
12.0 -- 14.0 & 12.94 & --                      & --                       & --                    & 16.8$\pm\,0.7\,\pm\,$1.0       & 16.18$_{-1.15}^{+1.18}$(15.93) \\
13.5 -- 15.0 & 14.22 & --                      & --                       & 19.80$\pm\,1.20$      & --                             & 16.23$_{-1.12}^{+1.14}$(16.00) \\
14.0 -- 16.0 & 14.95 & --                      & --                       & --                    & 17.9$\pm\,1.2\,\pm\,$1.3       & 16.26$_{-1.10}^{+1.12}$(16.04) \\
15.0 -- 17.0 & 15.95 & --                      & --                       & 20.80$\pm\,1.20$      & --                             & 16.30$_{-1.08}^{+1.10}$(16.09) \\
16.0 -- 18.0 & 16.96 & --                      & --                       & --                    & 18.3$\pm\,1.7\,\pm\,$1.2       & 16.33$_{-1.06}^{+1.08}$(16.13) \\
17.0 -- 20.0 & 18.40 & --                      & --                       & 22.00$\pm\,1.30$      & --                             & 16.38$_{-1.03}^{+1.05}$(16.18) \\
18.0 -- 20.0 & 18.96 & --                      & --                       & --                    & 19.8$\pm\,1.9\,\pm\,$1.3       & 16.39$_{-1.02}^{+1.04}$(16.20) \\
20.0 -- 25.0 & 22.28 & --                      & --                       & 24.50$\pm\,1.80$      & --                             & 16.47$_{-0.97}^{+0.98}$(16.30) \\
20.0 -- 25.0 & 22.29 & --                      & --                       & --                    & 19.5$\pm\,1.7\,\pm\,$1.3       & 16.47$_{-0.97}^{+0.98}$(16.30) \\
25.0 -- 30.0 & 27.31 & --                      & --                       & 18.10$_{-4.0}^{+3.3}$ & --                             & 16.56$_{-0.92}^{+0.93}$(16.40) \\
25.0 -- 30.0 & 27.33 & --                      & --                       & --                    & 23.6$_{-2.9}^{+2.6}$$\pm\,$1.6 & 16.56$_{-0.92}^{+0.93}$(16.40) \\
30.0 -- 40.0 & 34.36 & --                      & --                       & 28.50$_{-4.5}^{+3.9}$ & --                             & 16.64$_{-0.87}^{+0.87}$(16.49) \\
30.0 -- 40.0 & 34.46 & --                      & --                       & --                    & 18.8$_{-4.3}^{+3.5}$$\pm\,$1.3 & 16.64$_{-0.87}^{+0.87}$(16.50) \\
\end{tabular}
\end{ruledtabular}
\end{center}
\end{table*}
%%%%%%%%%%%%%%%%%%%%%%%%%%%%%%%%%%%%%%%%%%%%%%%%%%%%%%%%%%%%%%%%%%%%%%% Table III - End

It is instructive to make some important remarks concerning
the CELLO data reported in \cite{Behrend:1990sr}.
These data were presented for the quantity
$\frac{F^{2}M^3}{64\pi}~\text{eV} \equiv a$,
evaluated at the reference momentum scale
$\langle Q^2 \rangle \equiv \tilde{Q}^2$.
They have been converted here to the quantity $Q^2F(Q^2)$ using
the relation
$
 Q^2|F^{\gamma^*\gamma\pi^0}(\tilde{Q}^2)|
=
 \frac{1}{4\pi\alpha}
 \sqrt{\frac{64\pi a}{M^{3}}} |Q^2|~\text{GeV}
$,
where $M\simeq 135$~MeV and $\alpha=1/137$.

It is worth noting that the CELLO data are usually shown for the
quantity
$Q^2F^{\gamma*\gamma\pi^0}(Q^2)$
not at the scale $\tilde{Q}^2$ but rather at the symmetric point of
each $Q^2$ interval, i.e., at the scale
$\bar{Q^2}=(Q_\text{max}^2 + Q_\text{min}^2)/2$.
The resulting deviations of the scaled TFF
$\tilde{Q}^2F^{\gamma^{*}\gamma\pi^0}(\bar{Q}^2)$
from $\tilde{Q}^2F^{\gamma^{*}\gamma\pi^0}(\tilde{Q}^2)$
are very small at lower $Q^2$ but they increase
with $Q^2$, becoming strongest at the highest scale probed, viz.,
$\bar{Q^2}=2.40$~GeV$^2$ for which one has
$
 Q^2F^{\gamma^{*}\gamma\pi^0}(\bar{Q^2})[0.01 \times \text{GeV}]
=
 18.17_{-3.98}^{+3.25}
$
instead of
$
 Q^2F^{\gamma^{*}\gamma\pi^0}(\tilde{Q}^2)[0.01 \times \text{GeV}]
=
 16.43_{-3.60}^{+2.94}
$,
see Table \ref{tab:ff-values-table}.
\end{appendix}

\end{document}